\documentclass[reqno]{amsart}
\usepackage{graphicx,amsmath}
\usepackage{booktabs} 
\usepackage[normalem]{ulem}
\usepackage{t1enc}              
\usepackage[utf8]{inputenc}    
\usepackage{xcolor,hyperref}

\def\re{e}   
\def\fe{\mathsf{e}} 
\def\feb{\mathsf{u}}     
\def\ref{u_{f}}   
\def\ebflu{q} 
\def\fs{s} 

\def\fv{\mathsf{v}} 
\def\Bhopot{\Phi} 
\newcommand{\qint}{A}  
\def\LFden{\Gamma_{\! \rho}} 
\def\LFgden{\Gamma_{\!\!\nabla\!\rho}} 
\def\LFvel{\left.\Gamma_{\!\! v}\right.} 
\def\LFene{\Gamma_{\! e}} 
\def\eT{e} 
\def\qT{q_T} 

\def\qie{q} 

\newcount\n  \newdimen\w

\def\Repeat#1#2{\n=#1\relax\loop\ifnum       
  \n>0\relax #2\advance\n by-1\repeat}

\long\def\OMIT#1{\relax }  

\def\re#1{(\ref{#1})}   
  
\def\eqn#1#2{ \begin{align} \label{#1}         #2 \end{align}}

\def\nl#1{          \\ \label{#1}        }  
\def\nnl#1{ \tag*{} \\ \label{#1}        }  

\def\delim#1#2#3{\csname\ifcase#1 relax\or   
   big\or Big\or bigg\or Bigg\fi\endcsname   
  {\ifcase#2\or\Delim#3\or\deliM#3\fi}}      
\def\Delim#1{\ifcase#1\relax\or(\or[\or\{\or<\or\langle\or|\or\|\or---{ }\fi}
\def\deliM#1{\ifcase#1\relax\or)\or]\or\}\or>\or\rangle\or|\or\|\or{ }---\fi}

\let\f\frac                     
        
\def\largerfrac#1#2#3{      
  \whichtypesize\n=\currenttypesize\advance\n by #1 \mathchoice
  {\setbox0\hbox{$\displaystyle-$} \w=.5\ht0\advance\w by-.5\dp0\setbox0
    \hbox{\typesize\n $\displaystyle-$} \advance\w by -.5\ht0\advance\w
    by .5\dp0\raise\w \hbox{\typesize\n$\displaystyle{\frac{#2}{#3}}$}}
  {\setbox0\hbox{$-$} \w=.5\ht0 \advance\w by -.5\dp0 \setbox0\hbox
    {\typesize\n $-$} \advance\w by-.5\ht0\advance\w by
    .5\dp0\raise\w\hbox{\typesize\n$\frac{#2}{#3}$}}
  {\setbox0\hbox{$\scriptstyle-$} \w=.5\ht0 \advance\w by-.5\dp0\setbox0
    \hbox{\typesize\n $\scriptstyle-$} \advance\w by -.5\ht0 \advance\w
    by .5\dp0 \raise\w\hbox{\typesize\n$\scriptstyle{\frac{#2}{#3}}$}}
  {\setbox0\hbox{$\scriptscriptstyle-$} \w=.5\ht0
    \advance\w by -.5\dp0 \setbox0\hbox{\typesize\n
    $\scriptscriptstyle-$} \advance\w by -.5\ht0 \advance\w by .5\dp0
    \raise\w\hbox{\typesize\n$\scriptscriptstyle{\frac{#2}{#3}}$}}  }

\def\abs#1{{\left| #1 \right|}}

\def\d{{\rm d}}       

\begin{document}
\pagestyle{plain}  

\title{Holographic fluids: a thermodynamic road to quantum physics}
\author{{\large P. V\'an$^{1,2,3}$}\\ \\\tiny
$^1$Department of Theoretical Physics, Wigner Research Centre for Physics, H-1525 Budapest, Konkoly Thege Miklós u. 29-33., Hungary; \\
$^2$Department of Energy Engineering, Faculty of Mechanical Engineering,  Budapest University of Technology and Economics, H-1111 Budapest, Műegyetem rkp. 3., Hungary
\\$^3$MONTAVID Thermodynamic Research Group\\
email: van.peter@wigner.hu}

\date{\today}

\begin{abstract}
Quantum mechanics, superfluids, and capillary fluids are closely related: it is thermodynamics that links them. In this paper, the Liu procedure is used to analyze the thermodynamic requirements. A comparison with the traditional method of divergence separation highlights the role of spacetime. It is shown that perfect Korteweg fluids are holographic. The conditions under which a complex field can represent the density and velocity fields of the fluid, and where the complex scalar field becomes a wave function of quantum mechanics, are explored. The bridge between the field and particle representations of a physical system is holography, and the key to holography is the Second Law of thermodynamics.
\end{abstract}
\maketitle

{\em "A theory is the more impressive the greater the simplicity of its premises is, the more different kinds of things it relates, and the more extended is its area of applicability." in Albert Einstein: Autobiographical Notes}

\section{Introduction}

\subsection{Classical holography}
Holography is an expected property in theories of quantum gravity, which states that the volumetric forces on a test mass are equivalent to a lower-dimensional formulation on the boundary of the corresponding region \cite{tHo93m,Bek03a}. Holography is a principle motivated by black hole thermodynamics. Therefore, in general, a relation to thermodynamics is expected \cite{Bou02a}. In particular, the concept of entropic force gives a direct connection, where the holographic principle, together with the Unruh effect, leads to gravity \cite{Ver11a}. For Newtonian gravity, the holographic property and Poisson equation are closely related, and the Unruh effect does not play a role \cite{Hos10m}. 

It is easy to see that the Poisson equation implies holography: the force on a test mass induced by gravitational pressure on a closed surface is equivalent to the bulk gravitational force density integrated over the volume with nonzero mass density. Holography is a local identity here, without a system of screens, because the force density of a gravitational field, density multiplied by the gradient of the gravitational potential, can also be written as the divergence of a second-order tensor, the pressure of classical gravity:
\eqn{Ngrav_holo}{
\nabla\cdot \pmb{P}_{grav} = \rho\nabla\phi.
}
Here, $\phi$ is the gravitational potential in Newtonian theory, $\rho$ is the mass density, and $\pmb{P}_{grav} = \frac{1}{8\pi G}\left(\nabla\phi\cdot\nabla\phi\pmb{I} - 2\nabla\phi\nabla\phi\right)$ is the pressure of the gravitational field, where $\rho$ is the density, $\pmb I$ is the second-order identity tensor, $G$ is the gravitational constant, and the central dot denotes contraction. 

Moreover, thermodynamic principles extend and explain the above relation between the holographic property and the field equation. The Poisson equation of Newtonian gravity follows from thermodynamic principles when those principles are applied to a classical scalar field. Both the holographic pressure--force relation [Eq.~\re{Ngrav_holo}] and the Poisson equation emerge in the marginal case of zero dissipation for perfect fluids \cite{VanAbe22a,AbeVan22a,SzuVan22m}. 

In fluid mechanics, it is well known that ideal Euler fluids are holographic. This fact is a consequence of elementary thermodynamics applied to perfect fluids, in which the pressure tensor is the thermostatic scalar pressure: $\pmb{P} = p\pmb{I}$ and the following identities are applied
\eqn{Eflu_holo}{
	\nabla p = \rho (\nabla h-T\nabla s) = \rho (\nabla \mu+ s \nabla T),
}
where $h$ and $s$ are the specific enthalpy and specific entropy, respectively, $\mu$ is the chemical potential, and $T$ is the temperature. Therefore, in Euler fluids, the divergence of the pressure is also a force density, with the specific enthalpy as the potential in the case of isentropic processes (or barotropic fluids) and the chemical (Gibbs) potential in the case of isothermal fluids. 

One may wonder about possible generalizations. Therefore, it is worth clarifying the concept itself. In the following, a field theory, or a continuum, is called {\em classical holographic} if the following identity is valid for the second-order constitutive tensor field $P^{j}_{i}$ and for the constitutive scalar field $\Phi$:
\eqn{Gen_holo}{
	\partial_j P^{j}_{i} = \rho\partial_i\phi, \qquad \nabla\cdot \pmb{P} = \rho\nabla\phi.
}
A field is called constitutive if it is defined on the constitutive state space, spanned by the field variables and its derivatives. For example, for Euler fluids, the constitutive state space is the same as the thermodynamic state space and spanned by the specific entropy and the specific volume of the fluid, $(s,\rm{v})$. Usual thermodynamic variable transformations can be applied; for example, the right side of Eq.~\re{Eflu_holo} shows a Legendre transformation. 
Remarkably, Eq.~\re{Gen_holo} is formulated for nonrelativistic field theories.

When combined with the balance of momentum, Eq.~\re{Eflu_holo} gives the Friedmann form of the Euler equation \cite{Fri22t}:
\eqn{Eflu_eqm}{
	\rho \pmb{\dot v} + \nabla p = -\rho\nabla\varphi, \quad \rightarrow \quad 
	\pmb{\dot v} = -\nabla(\varphi+h)+ T\nabla s = -\nabla(\varphi+\mu)+ s\nabla T.	
}
Here, the momentum balance of the fluid is given in the Lagrangian form, where the overdot denotes the comoving, substantial time derivative. Therefore, in the case of barotropic fluids, the last term on the right-hand side of Eq.~\re{Eflu_holo} is zero, and the holographic property implies that the partial differential equations of fluid motion are given in the form of a Newtonian equation of a point mass. However, the analogy is incomplete, because the potential is not fixed; it depends on the state space. If the density distribution is not given, Eq.~\re{Eflu_eqm} is coupled to the continuity equation and to the conservation of mass:
\eqn{conm_eqn}{
	\dot \rho + \rho\nabla\cdot\pmb{v} =0.
}
Furthermore, one cannot avoid dealing with velocity as a field. The construction is similar to the idea of pilot waves without quantum mechanics. 

\subsection{Quantum fluids}

The relation of quantum mechanics and hydrodynamics, the quantum-hydro correspondence, goes back to the reformulation of the Schrödinger equation, 
\eqn{Sch_eqn}{
	i\hbar \frac{\partial \psi}{\partial t} + \frac{\hbar^2}{2m}\Delta \psi -V(\pmb{x},t)\psi = 0,
}
that is an evolution equation of the wave function, a complex scalar field $\psi$. Here, $i$ is the imaginary unit, $m$ is the mass of a quantum particle, and $\hbar$ is the reduced Planck constant. The above equation can be transformed to a fluid form with the Madelung transformation, which separates the amplitude, $R$, and the phase, $S$, of the wave function:
\eqn{Mad_traf}{
	\psi = R e^{i\frac{m}{\hbar}S}.
}
The real and imaginary parts of Eq.~\re{Sch_eqn} then give the continuity equation and the Bernoulli equation of irrotational flow \cite{Mad26a}, with the density given by $\rho = R^2 = \left|\psi\right|^2$, and the velocity potential given by $\pmb{v}=\nabla S$. A fluid theory thus emerges with the balance of mass [Eq.~\re{conm_eqn}] and the balance of momentum, 
\eqn{qhidro}{
	\rho \pmb{ \dot v} + \nabla\cdot\pmb{P_Q} = \pmb{0},
}
as evolution equations. However, this is a very particular fluid because the pressure tensor has the following form \cite{Hol93b}:
\eqn{Qpres_eqn}{
	\pmb{P}_Q = -\left(\frac{\hbar}{2 m}\right)^2  \rho \nabla\frac{\nabla\rho}{\rho}. 
}
It is easy to see that this pressure is holographic in the above sense because its divergence will be
\eqn{qhol_eqn}{
	\nabla\cdot \pmb{P}_Q = \rho \nabla \phi_B = - \frac{\hbar^2}{4 m^2} \rho\nabla\left(\frac{\Delta \rho}{\rho} - \frac{\nabla\rho\cdot\nabla\rho}{2\rho^2}\right).
}
In Eq.~\re{qhol_eqn}, $\phi_B$ is the Bohm potential, which is usually written as 
\eqn{Bpot}{
	\phi_B=-\frac{\hbar^2}{2 m^2}  \frac{\Delta R}{R}, 
}
with $R =\sqrt{\rho} = \abs{\psi}$. Therefore, the holographic relation given by Eq.~\re{qhol_eqn} connects the Bohmian and pilot-wave interpretations of quantum mechanics for a particular fluid form where the pressure tensor is a nonlinear function of the density, the gradient of the density, and its second gradient. This correspondence between hydrodynamics and single-particle quantum mechanics is fascinating and has inspired several extensions (see, for example, Jammer \cite{Jam74b}, and also Jackiw and coworkers in quantum field theories \cite{BisEta03a,JacEta04a}). 

From a continuum point of view, the pressure in Eq.~\re{Qpres_eqn} represents a perfect Korteweg fluid. Korteweg fluids were proposed as extensions of the static theory of van der Waals capillarity \cite{Waa894a}, where the static part of the pressure tensor depends on both the first and second spatial derivatives of the density in an isotropic manner \cite{Kor901a}: 
\eqn{Kortp_eqn}{
	\pmb{P}_K(T,\rho,\nabla\rho,\nabla^2\rho) &= \left(p - \alpha \Delta\rho -\beta (\nabla\rho)^2\right)\pmb{I} - 
	\delta \nabla\rho\circ \nabla\rho - \gamma\nabla^2\rho, 	\nonumber\\	
	P^{ij}(T, \rho, \partial_i\rho,\partial_{ij}\rho) &= \left(p - \alpha \partial_k^{\;k}\rho -\beta (\partial_k\rho\partial^k\rho\right)\delta^{ij} - 	\delta \partial^i\rho\partial^j\rho - \gamma\partial^{ij}\rho.
}
Here, $\alpha$, $\beta$, $\gamma$, and $\delta$ are density- and temperature-dependent material parameters like the thermostatic pressure $p$. Equation~\re{Kortp_eqn} was also given in index notation with spatial indices $i,j,k = 1,2,3$. Identical upper-lower indices indicate summation. The coefficients in the above formula cannot have arbitrary values; the compatibility with the Second Law of thermodynamics restricts the functional forms. The thermodynamic analysis of Sobrino and, later on, the analogous but more general analysis of Dunn and Serrin \cite{Sob76a,DunSer85a} revealed that the pressure of a perfect Korteweg fluid can be expressed with the help of the specific Helmholtz free energy, $f(T,\rho,\nabla\rho)$, as 
\eqn{Sobp_eqn}{
	\pmb{P}_S &= \left(\rho^2 \partial_\rho f - \rho \nabla\cdot (\rho \partial_{\nabla\rho}f)\right)\pmb{I} - \nabla\rho \circ \partial_{\nabla\rho}f.
}
Here, the lower-indexed $\rho$ and $\nabla\rho$ are for partial derivatives such as $\partial_\rho f = \frac{\partial f}{\partial\rho}(T,\rho,\nabla\rho)$. The thermodynamically compatible Korteweg pressure has the holographic property as well because at constant temperature,
\eqn{Kort_holo}{
	\nabla\cdot \pmb{P}_{S} = \rho\nabla\phi_S = \rho\nabla\left(\partial_\rho(\rho f)- \nabla\cdot\partial_{\nabla\rho}(\rho f)\right).
}
Therefore, the functional derivative of the free energy density, $\rho f$, is the potential of the mechanical force field. This quantity is most frequently interpreted as the generalized, gradient-dependent chemical potential, $\mu_S = \frac{\delta \rho f}{\delta \rho}=\partial_\rho(\rho f)- \nabla\cdot\partial_{\nabla\rho}(\rho f)$. The holographic property of Korteweg fluids was recognized by several other authors, including Antanovskii \cite{Ant96a} and V\'an and F\"ul\"op \cite{VanFul06a}, and was incorporated into the variational principle-motivated phase-field approaches \cite{AndEta98a}. However, the pressure given in Eq.~\re{Qpres_eqn} cannot be written in Sobrino form [Eq.~\re{Kort_holo}]. Therefore, it is seemingly incompatible with thermodynamics and contradicts the Second Law. 

In quantum mechanics and quantum field theories, the complex field represents a probabilistic fluid, and the density field $\rho$ is a probability density of the position of a quantum particle with mass $m$. In capillarity phenomena, for Korteweg fluids, the density is a real mass density. The theory of macroscopic quantum systems, particularly superfluidity (and superconductivity) in the $\Psi$ theories of Ginzburg field equations, is both for complex and hydrodynamic fields (see, for example, Volovik \cite{Vol03b}, Khalatnikov \cite{Kha18b}, and Ginzburg \cite{Gin97a}). Both the quantum and hydro forms are useful and important, and $\rho = \abs{\Psi}^2$ represents a real mass (or charge) density. The related Gross--Pitaevskii, Ginzburg--Sobyanin, and logarithmic Bialynicki--Birula--Mycielski equations and their combinations represent a family of nonlinear Schrödinger equations in the form
\eqn{GL_eqn}{
	i\hbar \frac{\partial \Psi}{\partial t} + \frac{\hbar^2}{2m}\Delta \Psi -\Phi_{sf}(\abs{\psi},\pmb{x},t)\Psi = 0,
}
where $\Phi_{sf}$ is a density-dependent potential function. The particular forms model the experimental properties of superfluid He around the $\lambda$ phase transition \cite{Gro61a,Pit61a,GinSob88a,BiaMyc76a,ScoZlo19a}. These fluids are holographic by definition, and the particle-like pilot-wave equation of motion is characterized by a combination of the Bohmian and superfluid potentials. 

Therefore, according to the above-mentioned examples, Euler fluids and perfect Korteweg fluids are holographic classical field theories connected to complex field representations of superfluids and the Madelung representation of quantum mechanics. The following scheme represents the usual relation:
$$
\boxed{\text{quantum mechanics}} \rightarrow 
\boxed{\text{superfluids}} \rightarrow
\boxed{\text{capillary fluids}},
$$
where the arrows are interpreted as analogies, thermodynamic compatibility is not an issue, and the holographic property looks accidental. 

As we have seen, for Newtonian gravity and for Euler and Korteweg fluids, the origin of the classical holographic property is connected to the Second Law of thermodynamics and to nonequilibrium thermodynamics. The relation to the Second Law is well expected in the case of simple Euler fluids in local equilibrium and can be understood through capillarity phenomena, in which a thermodynamic potential is also allowed to depend on the density gradient. The purpose of this paper is to analyze the reverse relation,
$$
\boxed{\text{capillary fluids}} \Longrightarrow
\boxed{\text{superfluids}} \Longrightarrow
\boxed{\text{quantum mechanics}},
$$
and classify the related conditions. We will see that from a mathematical point of view, the relations are not analogies: superfluids are particular Korteweg fluids, and Madelung fluids are particular superfluids. 

There are underlying conceptual questions: How can one connect thermodynamics to gravity and, more fundamentally, to quantum mechanics? Can a theory of individual particles (Schrödinger equation) and pure fields (gravity) be based on the emergent, collective behaviour of particulate matter? 

The fundamental balances and the Second Law inequality are universal and independent of the material's structure. The methodology of nonequilibrium thermodynamics is general and applicable to both particulate matter and fields. If the analysis aims to separate the universal and structure-dependent aspects, then thermodynamics is primary compared to statistical physics. Particular material structures (e.g., a material composed of rigid particles with vacuum interactions) introduce additional mathematical conditions that must be compatible with the more general universal considerations. One of the most remarkable consequences and a justification of the thermodynamic methodology is that the nondissipative part of the field equations emerges in a Euler--Lagrange form as functional derivatives without any variational principles \cite{VanKov20a}. 

The remainder of this paper is organized as follows. Section 2 treats the nonequilibrium thermodynamics of Korteweg fluids with the Liu procedure in an Eulerian laboratory frame of reference. The spacetime aspects are best shown in this way, and the covariance of the final form (i.e., the entropy production) is apparent. Section 3 presents the classical, heuristic treatment in a Lagrangian comoving frame via divergence separation, based on the Gibbs relation. A comparison of the divergence separation method and Liu procedure is instructive in understanding their proper usage and meaning. Section 4 discusses the previous results in the context of quantum mechanics and analyzes the conditions under which a complex field and a wave function can represent a Korteweg fluid. It is shown that the Bohm potential is either the consequence of the wave function representation or, at the same time, the consequence of separability of independent particles. That separability connects the multiplicative representation of probabilistic independence and additive representation of thermodynamic independence. Finally, Section 5 discusses additional aspects of the theoretical framework and some consequences. We argue that there is a novel thermodynamic method of quantisation that works without a Hamiltonian structure of the evolution equations and is applicable to dissipative evolution equations as well.

\section{Nonequilibrium thermodynamics of Korteweg fluids}

\subsection{Spacetime aspects: balances and thermodynamics}\label{matfraind}

The objectivity of classical continua is a long-discussed subject. The absolute time of Galilean relativistic spacetime separates space and time and four-dimensional aspects; four quantities are not apparent. Therefore, the mathematical representation of the objective and frame-independent formulation of nonrelativistic theories is discussed and controversial. Absolute time is best represented by an affine foliation of the four-dimensional affine space. Time cannot be an embedded submanifold. The nomination ``nonrelativistic'' is deceiving because it is relativistic in a Galilean sense. See Ref. \cite{Mat20b} for the precise definitions and mathematical background.  Correctly considering the spacetime aspects in extensions of classical nonequilibrium thermodynamics is crucial.

There are two aspects of objectivity: the problem of frame-independent physical quantities and the problem of constitutive relations. In both cases, realising that fundamental balances are four-divergences is helpful to identify four quantities and formulate objectivity requirements \cite{Rug89a,MusRes02a}; however, without a clear geometrical concept of Galilean relativistic spacetime, the transformation rule-based approach can be misleading \cite{MatVan06a}. The usual transformation rule-based definition of objectivity \cite{Nol67a} excludes momentum and velocity from the thermodynamic state space. This is a blocking problem because thermodynamics is best formulated in a comoving material frame with comoving quantities. Without a covariant approach, the thermodynamics of fields cannot be formulated in nonrelativistic spacetime. However, it is possible to obtain frame-independent results despite the frame-dependent framework if some rules are respected.

It is crucial to have the correct mathematical representation of the absolute time of Galilean relativistic spacetime; it cannot be embedded in the absolute four-dimensional spacetime \cite{Fri83b,Mat20b}. Let us emphasize again that fluids are frame independent, and their theories can be formulated in a frame-independent theory (e.g., the Gibbs relation of local equilibrium and the entropy production itself \cite{VanEta17a,Van17a}). The Gibbs relation is not invariant but covariant: relative velocity appears due to frame transformation formulas. However, the requirement for unusual mathematical tools can be avoided by considering some simple rules: 
\begin{enumerate}
	\item  Spacetime derivatives are Galilean four-covectors. Therefore, when changing reference frames, their timelike part transforms, while their spacelike part is invariant, \cite{ContraCov}. Therefore, the substantial, comoving time derivative is the transformed partial time derivative to a comoving reference frame, and the spatial derivative is invariant. Consequently, {\em spatial derivatives are objective and invariant}. If scalar densities and their spatial derivatives span the constitutive state space, the constitutive functions are objective. For Korteweg fluids, the density and its gradient are objective. 
	
	\item  Energy is not a scalar but a component of a higher-order tensor in the Galilean relativistic spacetime. The so-called total energy is the energy in a given reference frame, while internal energy, the difference between the total energy and the kinetic energy, is the comoving form of the energy, analogous to the well-known special relativistic rules (see, for example, Ref. \cite{LanLif59b}). In Galilean relativistic spacetime, the quadratic kinetic energy is part of the Galilean transformation rule \cite{Van17a}.  
	
	\item Balances are four-divergences. The Second Law of thermodynamics is a constrained inequality. If the constraints for the entropy inequality are balances or their spatial derivatives, the result will be objective. 
	
	\item Relative velocity is the spatial part of the four-velocity vector. The timelike part of nonrelativistic four-velocity is the constant number 1 because of the absolute time; therefore, the spatial derivatives of the relative velocity in an inertial frame are absolute. Hence, if objective four-velocity can be a state variable, a relative velocity with respect to an inertial reference frame can also be a state variable. The concept of Noll forbids velocity-dependent state spaces, while proper spacetime concepts do not. This is the most notable difference between our approaches. The laboratory frame calculations in this section and comoving frame calculations in Section 3 lead to the same objective, reference frame-independent result, \cite{aNoll}.
\end{enumerate}


In special relativistic thermodynamics, four quantities, including four-velocity, cannot be avoided, and the simple transition of nonrelativistic thermodynamic theory with comoving representation is problematic. For example, the paradox of temperature transformation refers precisely to the distinction of comoving and covariant concepts and the transformation properties of the energy (see, for example, Refs. \cite{Ein07a}--\cite{vKa68a}). The resolution of this paradox is based on the careful separation of the velocity concepts and the usage of spacetime quantities \cite{BirVan10a}. 

In the following calculations, the usual relative quantities are used. However, two independent derivations are presented: in the first, velocity is part of the constitutive state space; in the second, it is not, and Noll objectivity is respected. In addition, the two derivations use different methods of Second Law analysis. First, Korteweg fluids are represented in the Eulerian inertial reference frame, and the entropy production is calculated with the help of the Liu procedure. This way, the presentation of the thermodynamic methodology is more transparent. Next, Korteweg fluids are represented in a Lagrangian frame, and entropy production is calculated by divergence separation. Here, velocity is not part of the constitutive state space, and the rules of Noll objectivity are respected. In the second case, the calculation is more straightforward but more heuristic. In both cases, one obtains the same, apparently objective result. Both derivations are instructive. When interpreting and explaining the results, one can obtain a firm grasp of the methodology and clear insight into the interplay between the spacetime aspects and Second Law requirements. The constructive nature of the methodology is also demonstrated.  

\subsection{Balances and state spaces}
In this section, the formulas are presented both with and without indices. The index notation is best considered abstract; it does not refer to Descartes's coordinates but clarifies the tensorial properties and contractions of higher-order tensors. The upper and lower indices are distinguished, and identical ones denote contractions. 

The fundamental balances of mass, momentum, and energy are first given in a local form from the point of view of an external, inertial observer. The balance of mass is
\eqn{nlf_bmassloc}{
	\partial_t{\rho}+\partial_i(\rho v^i) = 0, \qquad 	\partial_t{\rho}+ \nabla\cdot(\rho \pmb{v}) = 0,
}
where $\partial_t$ is the partial time derivative, and $\pmb{v}$ is the barycentric velocity. The single-component fluid is comoving with the mass, without conductive part, there is no diffusion flux. The balance of momentum is given by the Cauchy equation:
\eqn{nlf_bmomloc}{
	\partial_t(\rho v^i)+ \partial_j\left({P}^{ij} + \rho v^iv^j\right) = 0^i, \qquad
	\partial_t(\rho \pmb{v})+ \nabla\cdot\left(\pmb{P} + \rho \pmb{v}\pmb{v}\right) = \pmb{0},
}
where $\rho v^i$ is the momentum density, and ${P}^{ij}$ is the pressure tensor, defined as the conductive current density of the momentum. The pressure is a constitutive quantity (a function of the constitutive state space). 

The balance of energy is
\eqn{nlf_balenerloc}{
	\partial_t(\rho \eT)+\partial_i(\qT^i+\rho \eT v^i) = 0, \qquad \partial_t{\rho \eT}+\nabla\cdot(\pmb{q}_T+ \rho \eT \pmb{v}) = 0,
}
where $\eT$ is the specific energy. Therefore, $\rho \eT$ is the energy density. One can see that the total current density of the energy in a laboratory frame, $\qT^i+\eT v^i$, is written as a sum of the conductive and convective current densities ($\qT^i$ and $\rho \eT v^i$, respectively). 

In the substantial form of balances the conductive current densities of the currents are eliminated. The overdot denotes the substantial time derivative of the corresponding physical quantity, defined as $\dot\empty = d_t = \partial_t + v^i\partial_i$. With comoving time derivatives and the mass balance, the velocity-dependent parts can be eliminated, and the balances are obtained as
\begin{eqnarray}
	\dot\rho + \rho \partial_i v^i &= 0, \qquad\quad \dot \rho + \rho\nabla\cdot {\bf v} =& 0, \label{balmasssub}\\
	\rho\dot v^i + \partial_jP^{ij} &= 0^i, \qquad
	\rho\dot {\bf v} + \nabla \cdot \pmb{P} =& \pmb{0}, \label{balmomsub}\\
	\rho \dot \eT + \partial_i \qT^i &= 0, \qquad
	\rho\dot \eT + \nabla \cdot \pmb{\qT} =& \!0.	\label{balensub}
\end{eqnarray}
It is worth emphasising that both the local and substantial balances are spacetime four-divergences. The balance of internal energy is obtained by subtracting the balance of kinetic energy from the total energy. The specific internal energy is $\feb = \eT- \frac{v^2}{2}$; therefore,   
\eqn{balintesub}{
	\rho \dot \feb + \partial_i \qie^i = - P^{ij}\partial_iv_j, \qquad
	\rho\dot \feb + \nabla \cdot \pmb{\qie} = - \pmb{P}:\nabla\pmb{v}.
}
Here, the conductive current density of the internal energy (i.e., the heat flux) is defined as $\qie^i = \qT^i- P^{ij}v_j$.

The closure of the system is complete if the constitutive functions (the pressure tensor $P^{ij}$ and the heat flux $\qie^i$) are given. They are restricted by the Second Law of thermodynamics and by material symmetries. The thermodynamic restrictions are determined with the help of the entropy balance. The entropy density $\rho s$ and the entropy flux $J^i$ are the conductive part of the local entropy current density and are constitutive quantities. The local and substantial forms of the entropy balance are 
\eqn{balentloc}{
	\partial_t (\rho s) + \partial_i(J^i + \rho s v^i) &\geq 0,\qquad
	\partial_t(\rho s) + \nabla \cdot (\pmb{J}+\rho s \pmb{v}) \geq  0, \\
	\rho \dot s + \partial_i J^i &\geq 0	, \qquad
	\rho\dot s + \nabla \cdot \pmb{J} \geq 0.	\label{balentsub}
}

The entropy balance is a conditional inequality that is subject to constraints determined by the field variables and the physical model of the continuum. In our case, the constraints are the simplest and most general ones; that is, they are the fundamental balances. 

In the following, we distinguish three kinds of state spaces. The constitutive functions, $\pmb{q}$, $\pmb{P}$, $s$, and $\pmb{J}$, are defined on the {\em constitutive state space (CSS)}. The primary constitutive function is the specific entropy (or the entropy density). The entropy inequality, the CSS, and the constraints determine the {\em thermodynamic state space (TSS)}, which is a reduction of the CSS, and the {\em process direction space (PDS)}, which is an extension of the CSS \cite{Mus90b}. For Korteweg fluids, the CSS is spanned by the specific internal energy, mass density, and their gradients, by the velocity and its gradient, and by the second spatial derivative of the density: ($e$, $\nabla e$, $\rho$, $\nabla\rho$, $\nabla^2\rho$, $\pmb{v}$, $\nabla\pmb{v}$). For Fourier--Navier--Stokes fluids, TSS is spanned by the specific internal energy and the mass density, $(e,\rho)$. For Korteweg fluids, the TSS is larger, and the specific entropy also depends on the density gradient, $(e,\rho,\nabla\rho)$. In the simplest cases, the PDS is spanned by the time and space derivatives of the fields in the constitutive state space that are not already in the constitutive state space. Hence, for Korteweg fluids, the PDS is spanned by $(\partial_t e, \partial_t\nabla e, \nabla^2 e, \partial_t \rho, \partial_t\nabla \rho, \partial_t \nabla^2 \rho, \nabla^3 \rho, \partial_t \pmb{v},  \partial_t\nabla\circ \pmb{v}, \nabla^2\pmb{v})$. 

The starting point of the analysis is the CSS, which is the domain of constitutive functions. The PDS and TSS are obtained through Second Law analysis. In the following sections, we calculate the thermodynamic requirements with the help of both the local and substantial forms using the Liu procedure and the more heuristic divergence separation. 

\section{Second Law-compatible Korteweg fluids: Liu procedure}

The inequality of the Second Law is obtained with the entropy balance [Eq.~\re{balentloc}] considering the physical conditions as constraints. In our case, these constraints are the fundamental balances [Eqs.~\re{nlf_bmassloc}, \re{nlf_bmomloc}, and \re{nlf_balenerloc}]. 

If the functional form of a constitutive function [e.g., the Korteweg pressure in Eq.~\re{Kortp_eqn}] is given, the Second Law can be used to test its thermodynamic compatibility. However, the Second Law can be used constructively, and one can obtain the most general form of the constitutive functions that is allowed considering the Second Law and the given constraints. The generality of the conditions determines the generality of the result. If the conditions are only the basic balances, the constitutive functions are universal, independent of material composition, and depend only on the representative capability of the constitutive state space. 

{The mathematical methods solve the conditional inequality with reducing that to an algebraic problem. In the Colemann-Noll procedure the constraints are substituted directly into the inequality, \cite{ColNol63a}. The Liu procedure combines the inequality and constraints with the help of multipliers, \cite{Liu72a}. The usage of multipliers is based on Farkas' lemma, a fundamental theorem of optimisation theory, \cite{Far894a,Far02a}. The connection of Liu procedure and Farkas' lemma was recognised by Hauser and Kirchner, \cite{HauKir02a}. Colemann-Noll and Liu procedures are equivalent, a detailed comparative study is \cite{MusEta01a}.}

For Korteweg fluids, for which the CSS is second-order weakly nonlocal in the mass density, the gradient of the mass balance [Eq.~\re{nlf_gbmass}] is a further constraint on the entropy inequality: 
\eqn{nlf_gbmass}{
	\partial_{it}{\rho}+\partial_{ij}(\rho v^j) = 0_{i}.
}

This is the most important speciality of the thermodynamic methodology in the case of weakly nonlocal state spaces \cite{Van05a,Cim07a}. The Liu procedure represents the Second Law inequality as a linear algebraic problem in the PDS. The PDS vectors can have any values; in particular, they can be both positive and negative depending on the related initial and boundary value problems. For example, Eq.~\re{nlf_gbmass} defines a linear algebraic condition of the $\partial_{it}\rho$ and $\partial_{ij}v$ PDS vectors. Further derivation of the constraint does not lead to more conditions because the higher derivatives are out of the PDS. Whether a constraint's derivative is a constraint depends on the CSS. Equation~\re{nlf_gbmass} is a constraint because the second gradient of the mass density is in the CSS.

With the $\LFden$, $\LFgden^{\;\;i}$, $\LFvel_i$, and $\LFene$ Lagrange--Farkas multipliers and the constraints given by Eqs.~\re{nlf_bmassloc}, \re{nlf_gbmass}, \re{nlf_bmomloc}, and \re{nlf_balenerloc}, the starting point of the Liu procedure is the following form of the entropy inequality:
\eqn{nlf_l1}{
	0  \leq\ &\partial_t (\rho s)+ \partial_i  J^i + \partial_i(\rho s v^i)-
	 \LFden\left[ \partial_t{\rho}+\partial_i(\rho v^i)\right]-
	\LFgden^{\;\;i}\left[ \partial_{it}{\rho}+\partial_{ij}(\rho v^j)\right]-
	\nnl{nlf_liu1}
	&\LFvel_i\left[\partial_t(\rho v^i)+ 
	\partial_j\left({P}^{ij} + \rho v^iv^j\right)  \right] -
	\LFene\left[ \partial_t(\rho e)+\partial_i(\qT^i+\rho e v^i)\right].
}

This formula gives
\begin{gather*}
	(s+\rho\partial_\rho {s}) \partial_t\rho + 
	\rho\partial_{\partial_i\rho} {s} \partial_{ti}\rho+
	\rho\partial_{\partial_{ij}\rho} {s} \partial_{tij}\rho+
	\rho\partial_{v^i} {s} \partial_{t}v^i+
	\rho\partial_{\partial_iv^j} {s} \partial_{tj}v^i+
	\rho\partial_e {s} \partial_te + \\
	\rho\partial_{\partial_ie} {s} \partial_{ti}e+\\
	+\partial_\rho {J}^i \partial_i\rho + 
	\partial_{\partial_j\rho} {J}^i \partial_{ij}\rho+
	\partial_{\partial_{jk}\rho} {J}^i \partial_{ijk}\rho+
	\partial_{v^j} {J}^i \partial_{i}v^j+
	\partial_{\partial_kv^j} {J}^i \partial_{jk}v^i+
	\partial_e {J}^i \partial_ie + \\
	\partial_{\partial_je} {J}^i \partial_{ij}e+ \nnl{nlf_l4}
	+\partial_i(\rho s v^i)-
	 \LFden\left( 
	\partial_t{\rho}+
	\rho\partial_i v^i+ 
	v^i\partial_i\rho\right)- \nnl{nlf_l7}
	- \LFgden^i\left( 
	\partial_{it}{\rho}+
	\rho \partial_{ij}v^j+
	v^j\partial_{ij}\rho+
	\partial_{i}\rho\partial_j v^j+
	\partial_{j}\rho\partial_i v^j\right)- \nnl{nlf_l8}
	- {\LFvel}_i\left(
	\rho\partial_t v^i+  v^i\partial_t\rho+ 
	\rho v^i\partial_jv^j+
	\rho v^j\partial_jv^i+
	v^j v^i\partial_j\rho+ 
	\partial_{\rho} {P}^{ij}\partial_{j}\rho+
	\partial_{\partial_k\rho} {P}^{ij}\partial_{jk}\rho+\right. \nnl{nlf_l9}
	\left. 
	+\partial_{\partial_{kl}\rho} {P}^{ij}\partial_{jkl}\rho+ 
	\partial_{v^k} {P}^{ij}\partial_{j}v^k+
	\partial_{\partial_{l}v^k} {P}^{ij}\partial_{jl}v^k +
	\partial_{e} {P}^{ij}\partial_{j}e+
	\partial_{\partial_ke} {P}^{ij}\partial_{jk}e \right) - \nnl{nlf_l10}
	- {\LFene}\left( 
	\rho\partial_t{e} + e\partial_t\rho +\rho e\partial_i v^i + 
	\rho v^i\partial_i e + e v^i\partial_i \rho +
	\partial_\rho {\qT}^i \partial_i\rho + 
	\partial_{\partial_j\rho} {\qT}^i \partial_{ji}\rho+\right. \nnl{nlf_l11}
	\left.  +\partial_{\partial_{jk}\rho} {\qT}^i \partial_{ijk}\rho+
	\partial_{v^j} {\qT}^i \partial_{i}v^j+
	\partial_{\partial_kv^j} {\qT}^i \partial_{ik}v^j+
	\partial_e {\qT}^i \partial_ie + 
	\partial_{\partial_je} {\qT}^i \partial_{ji}e\right)\geq 0.
\end{gather*}

Regrouping the terms according to the PDS components gives the inequality in the following form:
{
\begin{gather*}
	(s+\rho\partial_\rho {s} - \LFden- \LFvel_i v^{i}- \LFene e)\partial_t\rho + 
	 (\rho\partial_{\partial_i\rho} {s}-
	 \LFgden^i )\partial_{ti}\rho+
	 \rho\partial_{\partial_{ij}\rho} {s} \partial_{tij}\rho+
	\\+ (\rho\partial_{v^i} {s} -
	\rho \LFvel_i)\partial_{t}v^i+
	\rho\partial_{\partial_iv^j} {s} \partial_{tj}v^i+
	\rho(\partial_e {s}-
	 \LFene) \partial_te + 
	\rho\partial_{\partial_ie} {s} \partial_{ti}e+
	\\+(\partial_{\partial_{jk}\rho} {J}^i -
	 \LFvel_{l}\partial_{\partial_{jk}\rho} {P}^{li}-
	 \LFene\partial_{\partial_{jk}\rho} {\qT}^i+
	 \rho v^i\partial_{\partial_{jk}\rho} {s})\partial_{ijk}\rho+
	\\+(\partial_{\partial_kv^j} {J}^i -
	 \LFvel_{l}\partial_{\partial_{k}v^i} {P}^{lj}-
	 \LFene\partial_{\partial_kv^i} {\qT}^j+
	 \rho v^j \partial_{\partial_kv^i} {s})\partial_{jk}v^i+
	 \LFgden^i \rho \partial_{ij}v^j+
	\\+(\partial_{\partial_je} {J}^i - 
	 \LFvel_l\partial_{\partial_ie} {P}^{lj}+
	 \LFene\partial_{\partial_je} {\qT}^i+
	 \rho v^i \partial_{\partial_i e}s )\partial_{ij}e + 
	\\+\partial_\rho {J}^i \partial_i\rho + 
	 \partial_{\partial_j\rho} {J}^i \partial_{ij}\rho+
	 \partial_{v^j} {J}^i \partial_{i}v^j+
	 \partial_e {J}^i \partial_ie + 
	\\+\rho v^i\left(
	\partial_\rho {s}\partial_i\rho + 
	\partial_{\partial_j\rho} {s} \partial_{ji}\rho+
	\partial_{v^j} {s} \partial_{i}v^j+
	\partial_e {s} \partial_i e	\right)+ 
	 \\+ (s-\LFden - \LFvel_i v^{i} - \LFene e)\left( 
	\rho\partial_i v^i+ 
	v^i\partial_i\rho\right)-
	\\- \LFgden^i\left(
	v^j\partial_{ij}\rho+
	\partial_{i}\rho\partial_j v^j+
	\partial_{j}\rho\partial_i v^j\right)-
	\\- \LFvel_i\left(
	\rho v^j\partial_jv^i+
	\partial_{\rho} {P}^{ij}\partial_{j}\rho+
	\partial_{\partial_k\rho} {P}^{ij}\partial_{jk}\rho+
	\partial_{v^k} {P}^{ij}\partial_{j}v^k+
	\partial_{e} {P}^{ij}\partial_{j}e \right) -
	\\- \LFene\left( 
	\rho v^i\partial_i e+
	\partial_\rho {\qT}^i \partial_i\rho + 
	\partial_{\partial_j\rho} {\qT}^i \partial_{ji}\rho+
	\partial_{v^j} {\qT}^i \partial_{i}v^j+
	\partial_e {\qT}^i \partial_ie\right)\geq 0.
\end{gather*}
}
The coefficients of the PDS vectors must be zero, and one obtains the following Liu equations:
\eqn{fliu1}{
	\partial_t \rho \ &:\ & s+\rho\partial_\rho {s} - \LFden- \LFvel_i v^{i}- \LFene e &=0,\nl{fliu2}
	\partial_{ti} \rho &:\ & \rho\partial_{\partial_{i}\rho}   s-
	 \LFgden^i&=0^i,  \nl{fliu3}
	\partial_{tij} \rho &:\ & \partial_{\partial_{ij}\rho}   s &= 0^{ij},
	\nl{fliu4}
	\partial_t v^i \ &:\ & \partial_{v^i}  {s}- \LFvel_i &=0_i,\nl{fliu5}
	\partial_{tj} v^i &:\ & \partial_{\partial_{j}v^i}   s&=0^j_i,  
	\nl{fliu6} 
	\partial_t e \ &:\ & \partial_e  {s}- \LFene &=0,\nl{fliu7}
	\partial_{ti}e  &:\ & \partial_{\partial_{i}e}   s&=0^i.  
}
{
	Then, with Eqs. \re{fliu3}, \re{fliu5} and \re{fliu7}, one gets}
\eqn{fliu8}{
	\partial_{ijk} \rho &:\ & \partial_{\partial_{ ( kj}\rho}   J^{i )} &=
	 \LFvel_l\partial_{\partial_{ ( kj}\rho} {P}^{li )}+
	 \LFene \partial_{\partial_{ ( kj}\rho} {\qT}^{i )},\nl{fliu9}
	\partial_{jk}v^i  &:\ & \partial_{\partial_{ ( k}v^i}   J^{j )} &=
	 \LFvel_l\partial_{\partial_{ ( k}v^i} {P}^{lj )}+
	 \LFene \partial_{\partial_{ ( k}v^i} {\qT}_{j )}+
	\frac{\rho}{2} \LFgden^l(\delta_l^j\delta_i^k+\delta_l^k\delta_i^j),
	\nl{fliu10}
	\partial_{ij} e &:\ & \partial_{\partial_{ ( j}e}   J^{i )} &=
	 \LFvel_l\partial_{\partial_{ ( i}e} {P}^{lj )}+
	 \LFene \partial_{\partial_{ ( j}e} {\qT}^{i  )}. 
}

\noindent Here, the parentheses of the indices denote the symmetric part of the tensor (e.g., $A^{(ij)} = (A^{ij}+A^{ji})/2$). Due to Eqs.~\re{fliu3}, \re{fliu5}, and \re{fliu7}, the entropy density does not depend on the partial derivatives $\partial_{ij}\rho$, $\partial_{j}v^i$, and  $\partial_{i}e$. Equations~\re{fliu1}, \re{fliu2}, \re{fliu4}, and (\ref{fliu6}) connect the Lagrange--Farkas multipliers and entropy derivatives as follows:
\eqn{LFmult}{
	\rho\partial_\rho  {s} =  \LFden +  \LFvel_iv^i + \LFene e - s, \quad
	\quad
	\partial_{v^i}  {s} = \LFvel_i, \quad
	\partial_e  {s} = \LFene, \quad 
	\rho\partial_{\partial_{i}\rho}   s =  \LFgden^i.
} 

\noindent Note that the symmetry of the coefficients influences the Liu equations and should be explicitly considered in Eq.~\re{fliu9}. The solution of the system of Eqs.~\re{fliu8}--\re{fliu10} leads to the following entropy flux:
\eqn{LFcurr}{
	{J}^i =\  \partial_e {s} {\qT}^i+
	\frac{\rho^2}{2}\left(\partial_{\partial_i\rho}  s \partial_jv^j+
	\partial_{\partial_j\rho}  s \partial_jv^i\right) +
	\partial_{v^j}  s  {P}^{ji} + 
	\mathfrak{  J}^i(e,\rho,\partial_i\rho,v^i).
}

\noindent Here, the residual entropy flux $\mathfrak{  J}^i$ is not restricted; it can be an arbitrary function. The Liu equations are then completely solved, and the functional forms of the entropy density and entropy flux are restricted.  {
The final expression of the entropy production rate is simplified with the help of the restricted entropy density, $s=s(e,v^i,\rho,\partial_i\rho)$, the entropy flux, \re{LFcurr}, and the Lagrange-Farkas multipliers, \re{LFmult}, expressed by entropy derivatives. One obtains {\em the dissipation inequality}, in the following form}:
\begin{multline}
	0\leq \sigma_s = {\qT}^i\partial_i(\partial_e\fs)+
	P^{ij}\partial_i\left(\partial_{v^j} \fs\right)- \\
	-\partial_jv^j\left[
	\rho^2 \partial_\rho  {\fs}-
	\frac{\rho^{2}}{2}\partial_i\left(\partial_{
		\partial_i\rho}   \fs\right) \right]+
	\partial_jv^i\left[
	\frac{\rho^{2}}{2}\partial_i\left(\partial_{
		\partial_j\rho}  \fs\right)\right]\label{preeprod}
\end{multline}

Let us assume that the residual, local part of the entropy flux is zero, $\mathfrak{J}^i\equiv 0$, and  the internal energy is defined as the difference between the total and kinetic energies ($\feb :=e- v^2/2$). Both conditions are usual and also natural. The second condition is also justified by the expected covariance. Based on the partial derivatives of the entropy, the Lagrange--Farkas multipliers can be identified by thermodynamic intensives:

{
\eqn{def_tstat}{
	\frac{1}{T} := \partial_\feb \fs = \partial_e s, \quad 
	\frac{p}{T} := \partial_\fv \fs = -\rho^2 \partial_\rho \fs, \quad 
	\frac{\qint^i}{T}:= \partial_{\partial_i \rho}s, \quad 
	\frac{v^i}{T}:= -\partial_{v^i}s,
}
where $s = s(e-v^i/2,\rho,\partial_i\rho)$, $\fv=1/\rho$ is the specific volume, $T$ and $p$ are respectively the thermostatic temperature and pressure, and $A^i$ is a convenient notation to recover a traditional form of the Gibbs relation:
\eqn{Gibbs1}{
	d\fe- v_i dv^i = d\feb = Tds + \frac{p}{\rho^2} d\rho - \qint^i d \partial_i\rho.
}
The Lagrange-Farkas multipliers turn out to be the intensive state variables, defined by the derivatives of the specific entropy:  
\eqn{LF_tstat}{
	\LFene =  \frac{1}{T} =\frac{\partial \fs}{\partial \fe} =\frac{\partial \fs}{\partial \feb} , \quad 
	\LFvel_i =-\frac{v_i}{T} = \frac{\partial \fs}{\partial \feb}  \frac{\partial \feb}{\partial v^i}, \quad
	\LFgden^i = \rho\frac{\qint^i}{T}.
}

The multiplier of the mass balance is different, it can be expressed by the other derivatives and related to the chemical potential:
\eqn{LF_rho}{
	\LFden  = - \frac{ \feb-T \fs +\frac{p}{\rho}-v^2/2}{T} = - \frac{\mu_h-v^2/2}{T},
}
where the {\em homogeneous chemical potential} for the weakly nonlocal continuum is defined by the  homogeneous Gibbs--Duhem relation: $\mu_h := \feb-T \fs +\frac{p}{\rho}$. 

These are truly remarkable expressions indicating how entropy derivatives are connected to balances and how thermostatics emerges. The requirement of nonnegative entropy production enforces the static role of the entropy, connecting the fundamental balances to their multipliers. The first formula of \re{LF_tstat} identifies the multiplier of the energy balance by the reciprocal temperature. The second equality indicates, that the momentum balance is not a constraint in a comoving frame, and can be avoided introducing the internal energy, \cite{CimEta14a}. The third formula of \re{LF_tstat} determines a new intensive quantity. Finally, the Lagrange-Farkas multipliers of the mass balance in \re{LF_rho} is the Galilean transformed form of the chemical potential from a comoving to an external reference frame whose relative velocity with respect to the material is $v^i$. It is not the derivative of the specific entropy, it is the combination of the other derivatives, it is defined by the Euler relation. Both Euler relation and Gibbs relation are emerged, moreover, also the laboratory-comoving frame transformation rules of the thermodynamic quantities followed with Liu procedure. We can see that the Lagrange--Farkas multipliers of the local laboratory frame balances are the entropic intensive quantities in the laboratory frame, as expected.

Summarizing the second law restrictions for the specific entropy and the entropy flux, one obtains:
\eqn{nlf_s}{
	s &= \fs\left(\feb, \rho, \partial_i\rho\right), \nl{nlf_flux}
	{J}^i &=  \frac{1}{T}\left(\qie^i+
	\frac{\rho^{2}}{2}\left(\qint^i \partial_jv^j+\qint^j \partial_jv^i\right)\right).
}
}
Here, the conductive current density of the internal energy emerges as the difference between the total energy flux and the current density of the kinetic energy: \(\qie^{i}= \qT^i-v_j {P}^{ji}\). Finally, the dissipation inequality (i.e., the entropy production rate) is written as
\eqn{nlf_prod1}{
	0\leq \sigma_s = 
	\qie^i\partial_i\frac{1}{T}-\left(  P^{ij}-
	\left(p+ \frac{T\rho^{2}}{2}\partial_k\frac{\qint^k}{T}	\right)\delta^i_j
	- \frac{T\rho^{2}}{2}\partial_j\frac{\qint^i}{T}\right)\frac{\partial_iv^j}{T},
}
where the first part of the quadratic expression is the thermal part of the entropy production, and the second part is the mechanical part. The thermal part takes the usual form (i.e., the heat flux multiplied by the gradient of the reciprocal temperature). The second term is the product of the viscous pressure, the difference of the pressure tensor, $P^{ij}$, and the thermostatic pressure. If the entropy is independent of the density gradient, it simplifies to the usual scalar pressure, $p$. 

However, the thermal part of the entropy production is somehow incomplete. One may expect the separation of mechanical and thermal effects if the thermal part of the dissipation disappears when the entropy flux is zero. Therefore, we require that the entropy flux be parallel to the coefficient of the temperature gradient and rewrite the entropy production in the following form:
\eqn{nlf_prod}{
	0\leq \sigma_s = \left(\qie^i+\frac{\rho^{2}}{2}\left(\qint^i \partial_jv^j+\qint^j \partial_jv^i\right)\right)\partial_i\frac{1}{T}-
	\nnl{nlf_prod2}
	\left(  P^{i}_j-
	\left(p+ \frac{\rho^{2}}{2}\partial_k\qint^k\right)\delta^i_j
	- \frac{\rho^{2}}{2}\partial_j\qint^i\right)\frac{\partial_iv^j}{T}.
}

The final form of the entropy balance is
\eqn{nlf_sbalfin}{
	\partial_t(\rho s) + \partial_i\left(\frac{\qie^i- q^i_{ThK}}{T}+ \rho s v^i\right) = (\qie^i- q^i_{ThK}) \partial_i\frac{1}{T}-
	\left(  P^{i}_j- (P_{ThK})^{i}_j\right)\frac{\partial_iv^j}{T} \geq 0,
}
where 
\eqn{qkor_rev}{
	q^i_{ThK} = - \frac{\rho^{2}}{2}\left(\qint^i \partial_jv^j+\qint^j \partial_jv^i\right) \quad \text{and}\quad
	(P_{ThK})^{i}_j
	=\left(p+ \frac{\rho^{2}}{2}\partial_k\qint^k\right)\delta^i_j + \frac{\rho^{2}}{2}\partial_j\qint^i
}
are the Korteweg heat flux and Korteweg pressure tensor, respectively. If the heat flux $\qie^i$ and pressure tensor $P^{i}_j$ are equal to the Korteweg ones, the entropy production rate density is zero, and there is no dissipation. In this case, the Korteweg fluid is perfect. In the terminology of Dunn and Serrin, the Korteweg heat flux is the interstitial working; however, our expression here is different from theirs. We discuss the reason for this difference in the next section. The thermodynamic fluxes and forces are identified accordingly and given in Table \ref{tab_sff}.
\begin{table} \begin{tabular}{c||c|c}
		& Thermal & Mechanical \\\hline
		Force & $\partial_i \frac{1}{T} $ & $\frac{\partial_i v^j}{T}$ \\\hline
		Flux  & $\kappa^i = \qie^i+\frac{\rho^{2}}{2}\left(\qint^i \partial_jv^j+\qint^j \partial_jv^i\right)$  & 
		$\Pi^{i}_j=P^{i}_j- \left(p+ \frac{\rho^{2}}{2}\partial_k\qint^k\right)\delta^i_j
		- \frac{\rho^{2}}{2}\partial_j\qint^i$
	\end{tabular}\caption{\label{tab_sff} Adiabatic system of thermodynamic forces and fluxes.} 
\end{table}

Remarkably, the Korteweg pressure can then be written using the derivatives of the internal energy because, according to the Gibbs relation [Eq.~\re{Gibbs1}]:
\eqn{s_parc}{
	A^i = \left. \frac{\partial \feb}{\partial (\partial_i\rho)} \right |_{s,\rho},\qquad
	p  = \rho^2\left. \frac{\partial \feb}{\partial \rho} \right |_{s,\partial_i\rho}.
} 
Therefore, the mechanical part of the dissipation is connected to isentropic processes, as expected. 

The linear solution of the inequality assumes that thermodynamic fluxes are proportional to the thermodynamic forces; that is, the thermal flow and the viscous pressure are proportional to the gradient of the reciprocal temperature and the velocity gradient. For isotropic nonpolar materials with a symmetric pressure. that leads to four material parameters, the heat conduction coefficient ($\lambda$) and the bulk, shear, and rotational viscosities ($\eta_v$, $\eta$, and $\eta_r$, respectively):
\eqn{eq_onsrel}{
	\kappa^i =  \lambda \partial_i\frac{1}{T}, \qquad
	\Pi^{ij} = -\eta_v \partial_kv^k \delta^{ij} - \eta (\partial^i v^j+\partial^jv^i - 2\partial_kv^k\delta^{ij}/3) - 
	\eta_r (\partial^i v^j-\partial^jv^i).
} 
The linearization is analogous to the Fourier--Navier--Stokes system. 
 
\section{Thermodynamic constraints of Korteweg fluids: Divergence separation}

In this section, we derive the same results as in the preceding section using the simpler but more heuristic methodology of classical irreversible thermodynamics: the {\em divergence separation}. This method was introduced by Eckart and later used with the help of the hypothesis of local equilibrium \cite{Eck40a1,GroMaz62b}. The method is essentially a heuristic identification of the entropy flux and breaks up the bulk and surface terms of the entropy balance. In this section, the abstract index notation of the previous section are changed to the usual invariant notation of fluid mechanics and continuum mechanics with bold letters indicating vectors and tensors and nabla for spatial derivation. 

The first step is to determine the thermodynamic state space (TSS), the variables of the entropy density, or, as in the previous section, the variables of the specific entropy function. The hypothesis of local equilibrium is best expressed by the Gibbs relation of specific quantities. In our case, with gradients in the TSS, it is better called the {\em hypothesis of weakly nonlocal equilibrium}. We have already introduced the related Gibbs relation in Eq.~\re{Gibbs1}, with a convenient representation of the partial derivatives. {
However, the final result of the previous section is the starting point here.} With the nabla notation, this formula is written as
\eqn{nlf_Gibbsspec}{
	\d\feb=T\d \fs+\f{p}{\rho^2}\d\rho-\pmb{A}\cdot\d\nabla\rho.
}

A straightforward consequence is the Gibbs relation for the densities of internal energy and entropy:
\eqn{nlf_Gibbdens}{
	\d(\rho\feb)=T\d(\rho \fs) + \mu_h\d\rho - \rho\pmb{A}\cdot\d\nabla\rho,
}
where $\mu_h = e+p/\rho-T s$ is the homogeneous chemical potential. A complete homogeneous relation requires the density scaling of the gradient term as well:
\eqn{nlf_Gibbsgdens}{
	\d(\rho\feb)=T\d(\rho \fs) +\mu \d\rho -\pmb{A}\cdot\d(\rho\nabla\rho).
}
Here, $ \mu= \mu_h +\pmb{A}\cdot\nabla\rho$ according to the complete Euler homogeneity of the thermodynamic potential. Apparently, integrated extensive quantities of a corresponding homogeneous thermodynamic body with a homogeneous gradient leads to shape dependence, therefore thermodynamic relations of integrated extensive quantities of thermodynamic bodies with finite volume are not well defined. While we do not pursue that direction, we remark that thermodynamic relations are best introduced locally \cite{BerVan17b}. Following from Eqs.~\re{nlf_Gibbsspec} and \re{nlf_Gibbsgdens}: 
\eqn{Apdef}{
	\mu_h = \left.\frac{\partial (\rho\feb)}{\partial \rho}\right|_{\rho s,\nabla\rho}, \qquad 
	\rho \pmb{A} =  -\left.\frac{\partial (\rho\feb)}{\partial \nabla\rho}\right|_{\rho s,\rho}.
}

With the method of divergence separation, the entropy balance is determined by calculating the time derivative of the entropy and then introducing the constraints directly. This method is most convenient with the help of substantial forms because then the momentum balance is not necessary, and only the balances of mass and internal energy [Eqs.~\re{balmasssub} and \re{balintesub}] are used. The material time derivative of the specific entropy then follows as
\eqn{nlf_s1ebal}{
	\rho\dot{\fs} &=
	\rho\f{\dot{\feb}}{T} - \f{p}{T\rho}\dot\rho + \rho\f{\pmb{A}}{T}\cdot\f{d}{dt}\nabla\rho = \nnl{n}
	&= -\f{\nabla\cdot\pmb{\ebflu} + \pmb{P}:\nabla\pmb{v}}{T}+\f{p}{T}\nabla\cdot\pmb{v} -
	\rho\frac{\pmb{A}}{T}\cdot \nabla(\rho\nabla\cdot\pmb{v}) -
	\rho\nabla\rho\cdot\nabla \pmb{v}\cdot\frac{\pmb{A}}{T} = \nnl{n0}
	&= -  \nabla\cdot\frac{\pmb{\ebflu}}{T} + \pmb{\ebflu}\cdot\nabla\frac{1}{T} -
	\frac{\nabla\pmb{v}}{T}:\left(\pmb{P} - p\pmb{I}\right) - 
	\f{\pmb{A}}{T}\cdot\nabla\f{\rho^2}{2} \nabla\cdot\pmb{v}-
	\nabla\f{\rho^2}{2} \cdot\nabla \pmb{v}\cdot\frac{\pmb{A}}{T} - \nnl{trick}
	&\qquad \qquad - \boxed{	\rho^2\frac{\pmb{A}}{T}\cdot \f{\nabla\nabla\cdot\pmb{v}+\nabla\cdot(\nabla\pmb{v})}{2} }= \nl{n01}
	&=-  \nabla\cdot\frac{\pmb{\ebflu}}{T} + \pmb{\ebflu}\cdot\nabla\frac{1}{T} -
	\frac{\nabla\pmb{v}}{T}:\left(\pmb{P} - p\pmb{I} + {\pmb{A}}\cdot \nabla\f{\rho^2}{2} \pmb{I}+
	\nabla\f{\rho^2}{2} \pmb{A} \right)- \nnl{n02}
	&\qquad - \nabla\cdot\left(	\frac{\rho^2}{2T}\left[\nabla\pmb{v} \cdot\pmb{A} + \pmb{A}\nabla\cdot\pmb{v} \right]\right)+
	\left(\frac{\rho^2}{2T}\left[\nabla\pmb{v} \cdot\pmb{A} + \pmb{A}\nabla\cdot\pmb{v} \right]\right)	\cdot\nabla\f{1}{T}	+\nnl{n03}
	&\qquad +\frac{\nabla\pmb{v}}{T}:\left(\nabla\cdot\left[\f{\rho^2}{2}\pmb{A}\right]\pmb{I} 
	+\nabla\left[\f{\rho^2}{2}\pmb{A}\right] \right) =\nnl{n21}
	&=	-\nabla\cdot\left(\frac{\pmb{\ebflu}}{T} + 
	\frac{\rho^2}{2T}\left[\nabla\pmb{v} \cdot\pmb{A} + \pmb{A}\nabla\cdot\pmb{v} \right]\right) + \nnl{n2}
	&\quad	+\left(\pmb{\ebflu}+\frac{\rho^2}{2}\left[\nabla\pmb{v} \cdot\pmb{A} + \pmb{A}\nabla\cdot\pmb{v} \right]\right)	\cdot\nabla\f{1}{T} - \nnl{n3}
	&\quad	- \f{\nabla\pmb{v}}{T}:\left[\pmb{P} -\left(p+ \frac{\rho^{2}}{2}\nabla\cdot \pmb{A}\right)\pmb{I} - \frac{\rho^{2}}{2}\nabla\pmb{A}\right] \geq 0. 
}

\noindent Here, $\pmb{I}$ denotes the second-order identity tensor; the following formula was also used:
\eqn{nlf_dnab}{
	\f{d}{dt}\nabla\rho= -\nabla\rho\cdot\nabla\pmb{v}
	-\nabla\big(\rho\nabla\cdot\pmb{v}\big).
}
One can also identify the entropy flux and the entropy production rate density, giving the entropy balance as follows:
\eqn{ebal_fin}{
	&\rho\dot s + \nabla\cdot\left(\f{\pmb{\ebflu}}{T} + 
	\frac{\rho^2}{2}\left[\nabla\pmb{v} \cdot\f{\pmb{A}}{T} + \f{\pmb{A}}{T}\nabla\cdot\pmb{v} \right]\right) \nnl{ebf2}
	& = \quad\nabla\f{1}{T}\cdot\left(\pmb{\ebflu} + \f{\rho^2}{2}\left[\nabla\pmb{v} \cdot\pmb{A} + 
	\pmb{A}\nabla\cdot\pmb{v}  \right]\right)-\nnl{ebf3}
	&\qquad\f{\nabla\pmb{v}}{T}:\left[\pmb{P}-
	\left(p+ \frac{\rho^{2}}{2}\nabla\cdot \pmb{A}\right)\pmb{I} - \f{\rho^{2}}{2}\nabla\pmb{A}\right] \geq 0. 
}
This formula is identical to the form obtained in Eq.~\re{nlf_sbalfin} with the help of the Liu procedure in the previous section. 

\begin{table}
	\centering
	\begin{tabular}{c|c|c}
		&Thermal  &   Mechanical \\ \hline
		Forces & $\nabla \left(\frac{1}{T}\right)$ & 
		$-\frac{\nabla \mathbf{v}}{T}$\\ \hline
		Fluxes & $\pmb{\ebflu} + \f{\rho^2}{2}\left[\nabla\pmb{v} \cdot\pmb{A} + \pmb{A}\nabla\cdot\pmb{v}  \right]$ & 
		$\pmb{P}- \left(p+ \frac{\rho^{2}}{2}\nabla\cdot \pmb{A}\right)\pmb{I} - \f{\rho^{2}}{2}\nabla\pmb{A}$ \\ 
	\end{tabular}\\
	\caption{Thermodynamic fluxes and forces of Korteweg fluids: thermal version.}
	\label{tab_sff2} 
\end{table}

The linear isotropic solution of the inequality gives the Korteweg version of the Fourier--Navier--Stokes system of equations, with reversible contributions of Korteweg pressure and Korteweg heat flux [Eq.~\re{qkor_rev}]. The complete form of the constitutive functions of viscous, heat-conducting Korteweg fluids is
\eqn{heat_diss}{
	\pmb{q}   &= -\f{\rho^2}{2}\left[\nabla\pmb{v} \cdot\pmb{A} + \pmb{A}\nabla\cdot\pmb{v}\right] + 
	\lambda \nabla\f{1}{T},\nl{bvis_diss}
	tr\pmb{P}/3 &= p+ \frac{2\rho^{2}}{3}\nabla\cdot \pmb{A} - \eta_v \nabla\cdot \pmb{v}, \nl{svis_diss1}
	\pmb{P^{0s}} & = \f{\pmb{P}+\pmb{P}^T}{2} - \f{tr\pmb{P}}{3}\pmb{I} = 
	\f{\rho^2}{2}(\nabla\pmb{A})^{0s} - 2\eta (\nabla\pmb{v})^{0s} = \nnl{svis_diss}
	&=\f{\rho^2}{4}\left(\nabla\pmb{A}+(\nabla\pmb{A})^T - \f{2\nabla\cdot\pmb{A}}{3}\pmb{I}\right) -
	\eta \left(\nabla\pmb{v}+(\nabla\pmb{v})^T - \f{2\nabla\cdot\pmb{v}}{3}\pmb{I}\right), \nl{rotv_diss}
	\pmb{P^{a}} & = \f{\pmb{P}-\pmb{P}^T}{2} = \f{\rho^2}{2}(\nabla\pmb{A})^{A} - 2\eta_r (\nabla\pmb{v})^{A} =	 
	\f{\rho^2}{4}\left(\nabla\pmb{A}-(\nabla\pmb{A})^T\right) - \eta_r \left(\nabla\pmb{v}-(\nabla\pmb{v})^T\right).
}
Here, the upper indices $\pmb{0s}$ and $\pmb{a}$ denote the traceless symmetric and antisymmetric parts of the corresponding second-order tensors. The heat conduction coefficient $\lambda$ is related to the Fourier heat conduction coefficient: $\lambda_F= \frac{\lambda}{T^2}$. The terms $\eta_v$, $\eta$, and $\eta_r$ are the bulk, shear, and rotational viscosities, respectively. The second law requires that the viscosities and the heat conduction coefficient be non-negative. 

There are no cross effects; the scalar, vector, deviatoric, and antisymmetric tensorial components are independent according to the representation theorems of isotropic tensors. Remarkably, zero dissipation does not require zero heat flux and shear stress in Korteweg fluids. One expects that the fluid pressure relaxes to the perfect Korteweg pressure, as for normal viscous fluids in equilibrium. The last equation is also remarkable: in the case of Navier--Stokes fluids, the rotational viscosity ensures that the antisymmetric part of the pressure relaxes to zero \cite{GroMaz62b}. In the case of Korteweg fluids, it is not necessary.

The Liu procedure in the previous section and the divergence separation method above lead to the same results. The two methods are compatible. 

\subsection{Holographic perfect fluids.} 
A Korteweg fluid is called perfect if its entropy production is zero because of the material properties. This happens if the heat flux, $\kappa^i$, and the viscous pressure, $\Pi^{ij}$, are zero. Therefore the thermodynamically compatible current density of the internal energy and the thermodynamically compatible pressure tensor of perfect Korteweg fluids are
\eqn{qkor_rev1}{
	\pmb{q} =\pmb{q}_{ThK} = - \frac{\rho^{2}}{2}\left(\pmb{\qint} \nabla\cdot \pmb{v} + (\nabla\pmb{v})\cdot\pmb{\qint}\right), \nl{pkor_rev}
	\pmb{P}=\pmb{P}_{ThK}
	=\left(p+ \frac{\rho^{2}}{2}\nabla\cdot\pmb{\qint}\right)\pmb{I} + \frac{\rho^{2}}{2}\nabla\pmb{\qint}.
}
If the entropy is independent of the density gradients, the pressure tensor reduces to the Pascal pressure of a Euler fluid, and the heat flux is zero:
\eqn{nlf_Ef}{
	\pmb{P}_{\rm Euler} = -T\rho^2 \partial_\rho   \fs(\feb,\rho) \pmb{I} = 
	\rho^2 \partial_\rho   \feb(\fs,\rho) \pmb{I} = p \pmb{I}, \qquad
	\pmb{q} = \pmb{0}.
}
Remarkably, the thermostatic Pascal pressure is a different function for isentropic, isoenergetic, and isothermal processes. In the last case, the partial derivative of the Helmholtz free energy is to be considered: $p(T,\rho) = \rho^2 \partial_\rho  f(T,\rho)$, where $f = \feb-T\fs$ is the specific free energy. 

Perfect Korteweg fluids are classically holographic in the sense of Eq.~\re{Gen_holo} if some thermodynamic conditions are fulfilled. When calculating the divergence of the Korteweg pressure, one can see that
\eqn{nlf_potcond}{
	\nabla \cdot \pmb{P}_{ThK} =\rho\nabla\left(\mu_h+ \nabla\cdot(\rho\pmb{\qint})\right)+\rho s\nabla T = 
	\rho\nabla (h  + \nabla\cdot(\rho\pmb{\qint})) -\rho T \nabla s.
}
\noindent  Here, both the nonlocal enthalpy $\Bhopot_h=h  + \nabla\cdot(\rho\pmb{\qint})$ and the nonlocal chemical potential $\Bhopot_\mu = \mu_h+ \nabla\cdot(\rho\pmb{\qint})$ are mechanical potentials with homogeneous entropy and temperature fields, respectively. In both cases, the potential is a functional derivative. For the nonlocal enthalpy, the ``Lagrangian'' is the internal energy density, $\rho \feb(s,\rho, \nabla\rho)$:
\eqn{nlf_qpres_gen1}{
	\Bhopot_h =h  + \nabla\cdot(\rho\pmb{\qint})
	= \left.\frac{\partial(\rho   \feb)}{\partial\rho}\right|_{\rho s,\nabla\rho} - \nabla\cdot\left.\frac{\partial(\rho   \feb)}{\partial\nabla\rho}\right|_{\rho s,\rho}
	=\left.\frac{\delta  (\rho \feb)}{\delta \rho}\right|_{\rho s},
}
\noindent where $h=\feb+p/\rho$ is the specific enthalpy. It follows from Eq.~\re{nlf_Gibbdens} that for the nonlocal chemical potential, the ``Lagrangian'' is the free energy density $\rho f(T,\rho,\nabla\rho) = \rho (\feb - T s) $:
\eqn{nlf_qpres_gen2}{
	\Bhopot_\mu = \mu_h  + \nabla\cdot(\rho\pmb{\qint})
	= \left.\frac{\partial(\rho f)}{\partial\rho}\right|_{T,\nabla\rho} - \nabla\cdot\left.\frac{\partial(\rho   f)}{\partial\nabla\rho}\right|_{T,\rho}
	=\left.\frac{\delta  (\rho f)}{\delta \rho}\right|_{T}.
}

In the case of local equilibrium, if the thermodynamic potentials are independent of the density gradient, Eq.~\re{nlf_potcond} reduces to Eq.~\re{Eflu_holo} for Euler fluids. The Euler--Lagrange form (the functional derivative) emerges independently of any variational principle. This is in sharp contrast with phase-field approaches to Korteweg fluids, in which functional derivatives are combined with fluid mechanics \cite{AntEta97a}; the above formula is a basic assumption. 

A straightforward consequence of classical holography is that the momentum balance of a Korteweg fluid is like an equation of motion of a point mass in a force field given by a scalar potential:
\eqn{nlf_balNew}{
	\rho\dot{\pmb{v}} + \nabla\cdot\pmb{P}_{ThK} = \pmb{0} \quad
	\Longleftrightarrow \quad\dot{\pmb{v}} = - \nabla\Bhopot.
}
However, the similarity can be deceptive because $\Bhopot$ is not a fixed function of spacetime. $\Bhopot$ is not an external potential; rather, it depends on the density distribution and its derivatives and is best seen as a characteristic and particular self-force of the fluid. $\Bhopot$ can be interpreted as a local force of a test mass, but the density and energy fields of the continuum determine the force field itself. The continuity equation is not the only coupling; the energy balance is coupled to the mechanical system of equations, which is also the case for the seemingly pure mechanical homoentropic processes of perfect fluids. This is because the conductive part of the energy current density is not zero; rather, it is given by Eq.~\re{qkor_rev1}.  

\subsection{On the uniqueness of entropy production and interstitial working}

As was already mentioned, the interstitial working of Dunn and Serrin, [Eq.~\re{Sobp_eqn}], is similar but not the same as the perfect Korteweg heat flux [Eq.~\re{qkor_rev1}]. (Remarkable, that Sobrino introduced the same concept, but not the nomination.). Their calculations are based on the Helmholtz free energy as the thermodynamic potential, and the entropy production is calculated implicitly. The difference in the formulas results from the different calculations of the constraints. If one does not consider the symmetry of the second derivative of the velocity field, as in Eq.~\re{fliu9}, or the symmetric, boxed formula of Eq.~\re{trick}, only $\nabla\nabla\cdot\pmb{v}$ is used in the calculations, and one obtains the Korteweg pressure [Eq.~\re{Sobp_eqn}] along with the corresponding interstitial working. This mistake is best seen in Eq.~\re{Sobp_eqn} in Sobrino \cite{Sob76a} but is less apparent in Dunn and Serrin's elasticity-motivated finite deformation calculation. For homothermal or homoentrophic processes of perfect Korteweg fluids, the difference in the pressure term is a total divergence; therefore, one recovers the same holographic properties. 

Another aspect of interstitial working is the difference between the entropy production rate densities [Eqs.~\re{nlf_prod1} and \re{nlf_prod2}]. In the first formula, a heat flux, which is the multiplier of the gradient of the reciprocal temperature in the thermal term, is the current density of the internal energy that is missing in the interstitial working. In the second formula, the heat flux is parallel to the entropy flux. In the first case, the pressure of a perfect fluid depends on the temperature and temperature gradient; thus, the thermal and mechanical processes cannot be separated. This is also reasonable from a physical point of view: since one expects the heat flux to be parallel to the entropy current density, the temperature gradient drives the thermal interaction. Remarkably, the same decision must be made whenever the thermal interaction is not zero. The simplest case is thermodiffusion: the heat flux parallel to the entropy flux is called the true heat flux in chemical engineering \cite{KjeBed08b,KjeEta10b}. We will analyze the consequences of apparent uncertainty in a forthcoming publication \cite{SzuVan23m}.

\section{Field or particle: superfluids and quantum mechanics} 
\label{nlf_SchM}

As mentioned in the introduction, quantum mechanics and quantum field theories, both nonrelativistic and relativistic, can be reformulated in a fluid form. The relation to Korteweg fluids is also known: because the Bohm potential defines a particular chemical potential, it is a particular Korteweg fluid that has a complex field formulation. The key is Madelung transformation. Starting from the Schrödinger, Klein--Gordon, or Dirac equations, one can obtain a fluid form. Both fluid and quantum mechanics are field theories in that both the wave function and density--velocity fields are defined in spacetime. Nevertheless, the Schrödinger equation is a theory of a point mass, whereas fluid mechanics in general is not. Under what conditions becomes a fluid theory a theory of a point mass? This section formulates an answer to this question.

The holographic property of isentropic perfect fluids is only a partial explanation. There are two additional characteristics of quantum mechanics that should be represented in the fluid formulation:
\begin{itemize}
	\item Quantum mechanics is a probabilistic theory.
	\item The field equations of independent particles are independent and are additively separated.
\end{itemize}

The probability density of a two-particle system is represented by $\rho(\pmb{x}_1,\pmb{x}_2,t)$. For independent particles, the probability density is the product of individual probability densities: $\rho(\pmb{x}_1,\pmb{x}_2,t) =\rho_1(\pmb{x}_1,t) \rho_2(\pmb{x}_2,t)$. For normal fluids, the density of a mixture is the sum of the component densities, and a multicomponent quantum fluid is not like an ordinary one. However, the energy of the quantum field is additively separated for individual particles. This is usually a requirement of the Hamiltonian of the multiparticle system. For fluid theory, separability must be a requirement for the density dependence of the mass-specific internal energy:
\eqn{addcond_loc}{
	\feb(\rho_1(\pmb{x}_1,t) \rho_2(\pmb{x}_2,t)) = \feb(\rho_1(\pmb{x}_1,t)) +\feb(\rho_2(\pmb{x}_2,t)).
}

If $\feb$ is continuously differentiable, there exists a unique solution of this functional equation, namely, $\feb(\rho) = k \ln(\rho)$, where $k$ is a constant \cite{Jay57a,Ren61a}. This is fundamental to the relationship between statistical mechanics and equilibrium thermodynamics. In statistical mechanics, the additivity of thermodynamic entropy is the requirement; in quantum mechanics, the requirement is the additivity of evolution equations. We see that the additivity of the isentropic weakly nonlocal specific internal energy is our best choice to formulate a similar requirement. {
Since we require the Korteweg fluid to be isotropic, it depends only on the magnitude of the density gradient vector, $(\nabla\rho)^2$. For a two component fluid, the total spatial derivative of a two particle function will be denoted by $D = (\nabla_1,\nabla_2)$. Then the gradient of a multiparticle density is $D\rho(\pmb{x}_1, \pmb{x}_2,t) = (\nabla_{1}\rho, \nabla_{2}\rho)(\pmb{x}_1, \pmb{x}_2,t)$. Therefore, the additivity requirement can be formulated as}
\eqn{addcond_nloc}{
	\feb(\rho,(\nabla\rho)^2) = 
	\feb(\rho_1\rho_2,(\rho_2\nabla_1\rho_1)^2+(\rho_1\nabla_2\rho_2)^2) = 
	\feb(\rho_1,(\nabla_1\rho_1)^2) +\feb(s,\rho_2,(\nabla_2\rho_2)^2).
}
It is easy to see that a solution to this functional equation is 
\eqn{addsol}{
	\feb(\rho,(\nabla\rho)^2) = k \ln \rho + \frac{K}{2} \frac{(\nabla \rho)^2}{\rho^2},
}
where $k$ and $K$ are constants. More remarkably, the above solution is unique among continuously differentiable functions pu to an additive constant, \cite{VanFul06a}. If one expects that the above formula is the specific internal energy (i.e., the energy per unit mass), then \re{addsol} cannot change if $\rho$ is rescaled. Therefore, the logarithmic term cannot represent the energy density of a particle with mass $m$. In the following we introduce a slightly more general form of the specific internal energy of the Korteweg fluid, where the additive, gradient-dependent term, $\feb_F$, is separated from another term representing the usual local internal energy, $\feb_T$:
\eqn{inte_Fisher}{
	\feb_{qf}(s,\rho,\nabla\rho) = \feb_F+\feb_T = \frac{K}{2} \frac{(\nabla \rho)^2}{\rho^2 } + \feb_T(s,\rho).
}
We see that $\feb_{qf}$ is the thermodynamic potential of quantum fluids. According to the Gibbs relation [Eq.~\re{nlf_Gibbsspec}], one obtains the partial derivatives of the internal energy as
\eqn{qfint}{
	T= \frac{\partial \feb_T}{\partial s}, \quad 
	p= \rho^2\frac{\partial \feb_T}{\partial \rho} - K \frac{(\nabla \rho)^2}{\rho} , \quad
	\pmb{A} = K \frac{\nabla \rho}{\rho^2}.
}

The corresponding heat flux (interstitial working) and pressure tensor of a perfect Korteweg fluid are then
\eqn{qkor_qf}{
	\pmb{q}_{qf} &=  \frac{K}{2}\left(\nabla\rho \nabla\cdot \pmb{v} + (\nabla\pmb{v})\cdot\nabla\rho\right), \nl{pkor_qf}
	\pmb{P}_{qf} &=\left(\rho^2\left.\frac{\partial (\rho \feb_T)}{\partial \rho}\right|_s + \frac{K}{2}\Delta\rho\right)\pmb{I} +K\left(\frac{\nabla\rho\circ\nabla\rho}{\rho}-\frac{\nabla^2\rho}{2}\right).
}

The divergence of the pressure becomes
\eqn{hol_qf}{
	\nabla\cdot \pmb{P}_{qf} = \rho\nabla\left[\left.\frac{\partial (\rho \feb_T)}{\partial \rho}\right|_s + \frac{K}{2}\left(\frac{\nabla\rho\cdot\nabla\rho}{\rho^2}-2\frac{\Delta\rho}{\rho}\right)\right] - 
	\left.\frac{\partial (\rho \feb_T)}{\partial s}\right|_\rho \nabla s.
}
Here, the last term in the parentheses, which is the nonlocal part of the potential, can be written as 
\eqn{Bpot_eqn}{
	\frac{K}{2}\left(\frac{\nabla\rho\cdot\nabla\rho}{\rho^2}-2\frac{\Delta\rho}{\rho}\right) = 2K \frac{\Delta R}{R},
}
where $R=\sqrt{\rho}$. If $K= \frac{\hbar^2}{4m^2 }$. One can then identify the Bohm potential [Eq.~\re{Bpot}], which is the functional derivative of the internal energy density $\rho \feb_{qf} = \rho (\feb_F +\feb_T)$. More exactly, it is a partial functional derivative with constant specific entropy:
\eqn{var_Bpotspec}{
	\delta_\rho(\rho \feb_{qf})|_s =
	\left.\frac{\partial (\rho \feb_{qf})}{\partial \rho}\right|_{s,\nabla\rho} - \nabla\cdot\left.\frac{\partial (\rho \feb_F)}{\partial \nabla\rho}\right|_{\rho} =
	\left.\frac{\partial (\rho \feb_T)}{\partial \rho}\right|_s+ 2K \frac{\Delta R}{R}.
}
In addition, because
\eqn{pdid}{
	\left.\frac{\partial (\rho \feb)}{\partial \rho}\right|_{s,\nabla\rho} = \left.\frac{\partial (\rho \feb)}{\partial \rho}\right|_{\rho s,\nabla\rho} +Ts,
}
one obtains
\eqn{fund_sr}{
	\delta_\rho\feb(s,\rho,\nabla\rho)|_{\rho s} = 	\delta_\rho\feb(s,\rho,\nabla\rho)|_{s} - T s, 
}
because $s = \rho s/\rho$. Therefore, the functional derivative of the internal energy density can be written as
\eqn{var_Bpotdens}{
	\delta_\rho(\rho \feb_{qf})|_s = \delta_\rho(\rho \feb_{qf})|_{\rho s}  + T s = 
	\left.\frac{\partial (\rho \feb_T)}{\partial \rho}\right|_{\rho s}+ 2K \frac{\Delta R}{R} + Ts,
}
and the holographic property of the quantum fluid is expressed like that of a Euler fluid: 
\eqn{hol_qft}{
	\nabla\cdot \pmb{P}_{qf} = \rho\nabla\delta_\rho(\rho \feb_{qf})|_s - \rho T \nabla s =
	\rho\nabla\delta_\rho(\rho \feb_{qf})|_{\rho s} + \rho s \nabla T.
}
Therefore, quantum fluids are holographic in two particularly important situations: for homogeneous entropy or homogeneous temperature distributions (for homoentropic and homothermal materials). The respective potentials differ only in the thermal part; it is either the specific enthalpy or the chemical potential of the thermal part of the fluid equation of state.

\subsection{Wave function representation}

Starting from the field equations of quantum theory \cite{Mad26a,Tak57a}, the hydrodynamic, Bohmian, and pilot-wave formulations of quantum mechanics are derived from the Schrödinger, Klein--Gordon, and Dirac equations and also from field equations of quantum field theory \cite{JacEta04a}. For the example of the Schrödinger equation given in the introduction, the connection of quantum and fluid forms is based on the transformation of the variables by an appropriate version of the Madelung transformation:
\eqn{Mad0_traf}{
	\Psi = R e^{i\frac{S}{S_0}},
}

\noindent where $\rho = R^2$ is the probability density of the particle in a given position, and $S$ is the velocity potential ($\pmb{v} = \nabla S$). Substituting the amplitude--phase representation of the wave function, the formula \re{Mad0_traf}, into the quantum field evolution equations one can interpret the imaginary and the real parts separately. The imaginary part  results in the conservation of probability density, the continuity equation [Eq.~\re{conm_eqn}] and the real part gives the Bernoulli equation of the energy balance of a potential flow, with conserved vorticity. Therefore, the four-component classical field of $\rho,\pmb{v}$ can be represented by two scalar fields and the two scalar fields can be united into the well-known complex field equations. 

Starting from the quantum side, the continuum equations appear as mere ``interpretations,'' and hydrodynamics is an ``analogy'' because the Bohm potential or the quantum pressure define a very special fluid, and several aspects of the hydrodynamic form (e.g., the viscosity) are apparently not physical. In addition, if recognized, the classical holographic property looks \emph{ad hoc} and special. 

Starting from the fluid side, quantum equations are connected to clear conditions. If the momentum balance can be transformed to the Bernoulli form, it can also be represented by a complex scalar function. If a perfect, nondissipative Korteweg fluid is holographic and has a homogeneous temperature or homogeneous entropy field, there is a velocity potential for the gradient part of its velocity field ($\pmb{v} = \nabla S$), and one obtains the Bernoulli equation:
\eqn{Berncond}{
	\rho\pmb{\dot v} +\nabla\cdot\pmb{P}_K = 
	\rho\left( \partial_t\pmb{v} + \pmb{v}\cdot\nabla\pmb{v} +\nabla \Phi\right) =
	\rho\nabla\left(\partial_t S +\frac{\nabla S\cdot\nabla S}{2} + \Bhopot \right) = 0.
}
The equation follows from Eq.~\re{nlf_potcond} and because $\pmb{v}\cdot\nabla\pmb{v} = \nabla(\pmb{v}^2)/2$, if $\nabla\times\pmb{v} = 0$. The Bernoulli equation can then express the conserved specific energy (energy per unit mass) along a streamline:
\eqn{Bern}{
	\partial_t S +\frac{\nabla S\cdot\nabla S}{2} + \Phi(\rho,\nabla\rho) = const.
}

\noindent The continuity equation becomes
\eqn{conpsi}{
	\dot \rho +\rho\nabla\cdot \pmb{v} = \partial_t R^2 + \nabla(R^2\nabla S) =0.
}
Therefore, by multiplying Eq.~\re{conpsi} by $\frac{1}{2R}e^{i\frac{S}{S_0}}$ and Eq.~\re{Bern} by $\frac{iR}{S_0}e^{i\frac{S}{S_0}}$ and adding the formulas, one can separate the time derivative of the wave function as 
\eqn{preSchro}{
	\partial_t \psi + \frac{1}{2R}\left(2\nabla R\cdot \nabla S+ R\Delta S + i\frac{R}{S_0}(\nabla S)^2\right)\psi + \frac{i}{S_0}\Phi\psi =0.
}
Recognising the Laplacian of the wave function in the parentheses, one obtains
\eqn{preSchro2}{
	\partial_t \psi- \frac{iS_0}{2}\Delta\psi + \frac{i}{S_0}\left(\Phi + \frac{S_0^2}{2}\frac{\Delta R}{R}\right)\psi=0.
}

The natural unit of the velocity potential for a particle with mass $m$ is $S_0:= \frac{\hbar}{m}$. In addition, by multiplying the above equation by $i \hbar$, one obtains the evolution equation of a perfect Korteweg fluid in the form of a complex scalar field: 
\eqn{gSchro}{
	i\hbar	\partial_t \psi+ \frac{\hbar^2}{2m}\Delta\psi - m\left(\Phi + \frac{\hbar^2}{2m^2}\frac{\Delta R}{R}\right)\psi=0.
}

\noindent This is valid for any Korteweg potential $\Bhopot$ obtained in Eq.~\re{nlf_qpres_gen1} or \re{nlf_qpres_gen2}. Moreover, one can see that the Bohm potential form is separated naturally; therefore, if the gradient dependence comes from the additive energy form [Eq.~\re{inte_Fisher}], one can obtain the pure thermal part in the parentheses. However, for a quantum fluid, the parameter $K$ can depend on the entropy and spacetime; here, $S_0$ must be constant, or the wave function interpretation will be lost. 

Finally, it is worth formulating the quantum fluid form of the wave function evolution based on the internal energy in Eq.~\re{inte_Fisher} and $K=\frac{\hbar^2}{4 m^2}$ for homoentropic fluids:
\eqn{cSchro}{
	i\hbar	\partial_t \psi+ \frac{\hbar^2}{2m}\Delta\psi - m h(\rho,s)\psi=0.
}

\noindent From Eq.~\re{hol_qft}, one obtains $h=\left.\frac{\partial(\rho u_T)}{\partial \rho}\right|_s$, the specific thermal enthalpy. We have derived the complex field equation of superfluids, which is a generalisation of Ginzburg's $\Psi$ theory. If $h$ (or $\mu$) is a fixed field that is independent of the thermodynamic state variables, Eq.~\re{cSchro} is the Schrödinger equation of a particle with mass $m$. For quantum fluids, $m$ is a fixed parameter denoting the total mass of the fluid in the system and can be arbitrary, like in the case of superconductivity \cite{Gin97a}. 

\section{Summary}

Quantum mechanics, superfluids, and capillary fluids represent very different aspects of reality. In this paper, we have shown that the similarity in their theories is not accidental, it is based on a universal background that allows these systems to be treated uniformly in the framework of nonequilibrium thermodynamics. We have also shown that classical holography, the property that connects particles and fields, is a consequence of the Second Law of thermodynamics.

In classical holography, the surface and bulk forces are interchangeable because the divergence of the pressure equals a potential-generated force density. This particular holographic property is the consequence of the Second Law of thermodynamics in the marginal, nondissipative case of perfect fluids. Moreover, the potential has a particular form: it is a partial variational derivative of the internal energy density with respect to the density. The result was derived first with the rigorous Liu procedure and then with the help of divergence separation, with the heuristic and transparent method of classical irreversible thermodynamics. The latter derivation demonstrates that nonequilibrium thermodynamics can be extended from local equilibrium to {\em weakly nonlocal equilibrium}. 

Next, we explored the conditions of the complex scalar field representation. In this case, a general nonlinear Schrödinger-type equation emerges, where various models of superfluids (e.g., the Gross-Pitajevskii and Ginzburg--Soboyanhin equations) and the Schrödinger equation of a single particle are members of the Korteweg fluid family. The wave function representation is connected to a particular form of the internal energy with a natural probabilistic interpretation due to the unique additive form of the internal energy. 

The conditions that separate the different Korteweg fluids are enumerated below, with each point representing the next level of modelling conditions starting with the most general and becoming more special: 

\begin{enumerate}
	\item {\em Mass, momentum, and energy are conserved.} This is a universal condition that is independent of whether a particle-based or field-based representation of continua is used. 
	
	\item {\em Second Law, entropy balance.} One does not expect the Second Law to be directly applied to microscopic systems. The previous sections showed that the Second Law nontrivially restricts the equations of state of perfect fluids. A perfect, ideal system without dissipation is a marginal situation from the point of view of the Second Law. Moreover, according to our understanding of the atomic--nuclear--subnuclear hierarchical structure of matter, there is no reason to assume that there is no submicroscopic level below the Schrödinger equation.
	
	There is a remarkable, nontrivial aspect of the universality of Second Law based restrictions. We have seen that the obtained continuum equations are related to particular self-interacting fields. Such fluids are not well represented as set of particles and their vacuum fields. 
	
	\item {\em The fluid energy is sensitive to density inhomogeneities.} Therefore, our continuum is weakly nonlocal in mass density, and its energy depends on the spatial derivatives of the density. The holographic property is based only on these general conditions. For Korteweg fluids, the energy depends on first-order derivatives, while the pressure tensor and the equivalent potential field depend on second-order derivatives. Capillarity phenomena and surface tension follow from the static solution, assuming thermal and mechanical equilibrium. The definition of a perfect fluid also follows from the Second Law. 
	
	The presented methods, both Liu procedure and divergence separation, can be generalized to higher-order nonlocalities and also to other fields. An instructive example is a first-order, weakly nonlocal scalar field that turns out to be Newtonian gravity when Second Law restrictions are considered \cite{VanAbe22a,AbeVan22a}. 
	
	\item {\em The fluid is perfect.} The definition of perfect fluid comes from the requirement of conserved entropy when it is a consequence of material properties. For perfect fluids the Second Law is valid in a marginal case, the dissipation is zero. Perfect fluids are almost holographic. 
	
	\item {\em The entropy or the temperature is homogeneous.} This condition ensures that a perfect fluid is strictly holographic. Then mechanics can be separated from thermodynamics. If the thermal part of the evolution has not started or it is already finished, the mechanical potential in the momentum balance is the specific enthalpy or the specific Gibbs free energy, that is the chemical potential, respectively. 
	
	\item{\em The vorticity of the fluid is conserved.} In this case, there is a velocity potential, and therefore fluid fields can be represented by a complex scalar field, by a wave function. 
	\item{\em The Planck constant is constant.}  $K$ in Eq.~\re{addsol} is constant rather than being a state function. As a result, probabilistic separability is valid, and the nonlinear Schrödinger equation is obtained.
	
	\item{\em Density-independent potential.} Then the specific enthalpy, $h$, does not depend on the thermodynamic state variables; rather, it is a fixed function of time and position. 
\end{enumerate}

The first three conditions specify a thermodynamic system, namely, Korteweg fluids, a family of fluid theories that incorporates the Fourier--Navier-Stokes system of equations, among others. Korteweg fluids are a good model for capillarity phenomena and volcanic lava flow among others.

The fourth and fifth conditions mandate that the fluid is strictly holographic; therefore, the representation of the forces governing the motion of the fluid fields is dimensionally reduced. If the thermodynamic potential (the internal energy density $\rho\feb$) is independent of the density and depends only on spacetime, then one gets the equation of motion of a point mass. If the thermodynamic potential is density dependent, we obtain a pilot wave theory \cite{Bus15a,Fru22a}. This becomes a de Broglie--Bohm theory if the potential is the density gradient dependent Bohm potential.

The last three conditions separate various aspects of quantum systems. With the seventh condition, one obtains theories of superfluids, where the nonlinear Schrödinger equation represents the evolution of a wave function and the evolution of mass density. 

Finally, if all conditions are valid, one obtains the Schrödinger equation of a point mass, and the spacetime-dependent chemical potential becomes the potential energy when multiplied by the particle's mass. 

The above scheme of hierarchically arranged conditions outlines a uniform theoretical approach to various continuum and field theories. 

\section{Discussion}

\subsection{Universal thermodynamics}

Thermodynamics is considered a theory with limited validity. 
However, the presented results, that connect Korteweg fluids, superfluids and quantum mechanics, are based on the Second Law and could hardly be explained without it. Thermodynamics looks like more general than expected. 

Our fundamental assumption is that the Second Law is a universal condition, independent of material structure. Therefore, the Second Law agrees with any statistical model where the existence of entropy and the validity of the fundamental balances are part of the theory. The fact that gravity and quantum mechanics, the universal theories of physics, can be treated in this framework are the best arguments to reconsider the above-mentioned limited validity. 

Naturally, one may look for particular (and particulate) statistical models and explanations of the presented ideas because thermodynamics is an emergent theory. One may expect different microscopic mechanisms when modelling Korteweg fluids, superfluids, and quantum mechanics as emergent theories. Therefore, none of them can explain the uniform treatment. It seems impossible to find a {\em common} microscopic background for gravity on the cosmological and galactic scales and the quantum mechanics of an electron. What conceptual background can explain the uniform thermodynamic structure of these physical systems? From our perspective, the conceptual background is the stability of matter. The Second Law, and more specifically, the role of entropy, can be understood as theoretical tools to formulate evident and expected stability properties of equilibrium in any theory. It is easy to recognize the mathematical conditions of the Liapunov theorem when postulating entropy as a maximal and increasing quantity. A complete rigorous proof requires the detailed definition of the physical system; however, the analogy itself could be fruitful in mathematics \cite{FriLax71a} and insightful in physics \cite{GlaPri71b,Mat05b}. In our particular case, the form of entropy production indicates that the equilibrium is non homogeneous and defined by a perfect Korteweg fluid. The asymptotic stability of that equilibrium is a straightforward prediction from a physical point of view and a task to find boundary conditions and function spaces for rigorous mathematical statements. 

\subsection{Holographic principle.} The holographic principle is an expected property of quantum gravity and hydrodynamics \cite{KovEta03a} and also a source of inspiration, as in the case of AdS-CFT correspondence. The black hole origin indicates a deeper thermodynamic connection. For Newtonian gravity, holography is somehow trivial, but the formulation is simple and has a straightforward explanation: it is reduced to the observation that the gravitational force density can be transformed to a pressure tensor due to the field equation (the Poisson equation). The resulting concept is the {\em classical holographic property} [Eq.~\re{Gen_holo}]. As we have seen, classical holography is valid for perfect Korteweg fluids due to the Second Law, independent of any particular interaction. In the analogous parallel thermodynamic treatment, the Poisson equation is the perfect Newtonian gravity; therefore, the justification of classical holography is the same \cite{VanAbe22a}. 

The connection to thermodynamics is different from that of the entropic force concept of Verlinde \cite{Ver11a}, and its connection to quantum mechanics is independent of the concept of entanglement entropy (see Ref. \cite{NisEta06a1}). 

Any direct connection of our classical holography concept [Eq.~\re{Gen_holo}] to quantum gravity or string theory could be overly exaggerated; however, the logical relations are remarkable. In our case, the holographic property is not a condition but a consequence. Also in our case, the simple background comes with a direct and plausible interpretation. This interpretation is the critical aspect of the particle--field duality in quantum mechanics: quantum systems can be represented and modelled from both conceptual points of view, an aspect that is somehow overshadowed by the historical black hole origin.

Remarkable, that holographic considerations from quantum field theory, when applied to special relativistic or to Galilean relativistic hydrodynamics, are seemingly not related to  our concept of classical holography (see e.g. \cite{ArmJai20a,BagGou23a}). It is also remarkable that classical holography is not recognized in various recent generalisations of perfect fluids, \cite{BoeEta18a,BoeEta18a1}.  

\subsection{Dissipation and variational principles.} One does not expect the evolution equations of non-equilibrium thermodynamics to be derived from variational principles (contrary to, for example, Ref. \cite{diV22b}). The variational formulation is connected to perfect materials without dissipation. In our analysis, the mechanical potential that embodies the holographic principle has the form of a (partial) functional derivative where the ``Lagrangian'' is a thermodynamic potential. We emphasize that the variational form (the functional derivative) emerges without any variational principle as a consequence of the Second Law analysis. 

There are other ways to obtain similar results. For example, Pavelka \emph{et al.} applied a Poisson bracket structure \cite{PavEta18b}. In addition, given a functional derivative, one can find a suitable variational justification. However, the starting point is then an ideal system, and the dissipative part must be added to deal with the real world. This doubles the number of theoretical concepts and mathematical structures, and the uniform origin cannot be recognized. 

It is also remarkable that classical holography results in two kinds of dissipation. One type of dissipation is connected to the fluid flow (that is, to the linear solution of the entropy inequality) through viscosity and heat conduction. The second type of dissipation is connected to the streamlines [the Newtonian form of the evolution equation and the point mass representation in Eq.~\re{nlf_balNew}] through damping and friction. The damping term in the second case can be expressed as
\eqn{damp_hpw}{
	\dot{\pmb{v}} = - \nabla\Bhopot - \beta \pmb{v}.
}
The damping term is not a simple property if the potential $\Bhopot$ depends on the density or the gradient of the density, as in pilot-wave hydrodynamics. It can be a potential \cite{Bus15a,Fru22a} where the damping of the individual droplet motion is well expected and must be counterbalanced by the active excitation background. If we are in the Schrödinger level, then $\Bhopot$ is the Bohm potential, and Eq.~\re{damp_hpw} is equivalent to the Schrödinger--Langevin--Kostin equation \cite{Kos72a,Yas79a,SanSmi07a}, a dissipative version of the Schrödinger equation. A possible application of this remarkable idea is cooling with coherent control \cite{LosEta22a}.

\subsection{Quantum mechanics: interpretations and extensions.} Some remarks cannot be avoided regarding the foundations of quantum mechanics. As mentioned in the introduction, there is an enormous amount of literature on the various reformulations of quantum mechanics, including the hydrodynamic one. It is difficult to distinguish between scientific and speculative arguments. For example, the speculative completeness argument of the Copenhagen school disqualified the alternative approaches as mere ``interpretations.'' Therefore, no one is looking beyond quantum physics. Moreover, the connections between the interpretations have not really been analyzed. Approaches such as stochastic \cite{Fen52a,Nel66a}, pilot-wave, and Bohmian \cite{Hol93b,BohHil93b} approaches can reproduce quantum mechanics to some extent but also encounter various difficulties. One may wonder whether research on the connections between interpretations could be helpful when looking for new predictions.

Regarding the current analysis, from the general thermodynamic point of view, single-particle quantum mechanics is a very special Korteweg fluid in a large family of theories and models of various natural physical systems. In this case, the transitions are well defined and justified from a mathematical and physical point of view. Some known aspects of quantum mechanics appear from a new perspective. One of them is objectivity and frame independence, as discussed in the following subsection. In addition, some extensions of quantum mechanics are well motivated, like the logarithmic Schrödinger equation of Bialynicki-Birula and Mycielski \cite{BiaMyc76a} and a complex potential representing a mass source term \cite{Ben07a}, not to mention the various theories of superfluids. The concept of superfluidity at cosmological scales \cite{Hua16b} and the quark--gluon plasma \cite{AncEta19a,AncEta22a} are also somehow natural from the point of view of perfect Korteweg fluids. 

We remark that fluid mechanics is far from competing with the well-developed operator formalism and the related Copenhagen interpretation. Nevertheless, the simplicity of fluid models is surprising, as is the fact that Hilbert space operators and integration by projector measures can be substituted, even in a restricted sense, by fluid mechanics. 

Finally, the thermodynamic road to quantum mechanics is not related to a Hamiltonian structure of the evolution equations. It is also clear that it is a road and not a jump; one must respect the Second Law when investigating classical (e.g., Bohmian) mechanics and be consistent regarding the constraints of the classical evolution equations when introducing additive gradient energy contributions. The thermodynamic method is a novel and genuine approach to quantisation. One can test it with any classical continuum theories, including dissipative ones.

\subsection{Relativistic theory.} Our analysis is Galilean relativistic (nonrelativistic). As emphasized in Section \ref{matfraind}, the spacetime aspects of continuum theories are mostly hidden but essential. It is straightforward to prove that the final evolution equations of Korteweg fluids, which are also the dissipative ones, are not only Galilean covariants, but are also independent of reference frames. Comparing the Liu procedure and the divergence separation method highlights the hidden aspects of Galilean covariance when expressed with relative quantities. In addition, a complete Galilean covariant treatment is straightforward \cite{Van17a}. However, without further ado, the thermodynamic method cannot be generalized to special or general relativistic theories. Thermodynamics is based on the separation of space and timelike evolution, and this separation is based on the definition of comoving quantities, which requires the concept of the fluid's velocity field. This problem also appears with absolute time in Galilean relativistic theory \cite{Bre05a,OttEta09a,Van17a}. In the case of dissipative relativistic fluids, entropy production-based arguments are not sufficient, and instabilities appear \cite{HisLin85a, VanBir12a}. The application of the Liu procedure alone cannot clarify the problem \cite{Van09a}. 

However, starting from the quantum mechanical side, the fluid mechanical forms of the fundamental quantum mechanical evolution equations (e.g., the Klein--Gordon and Dirac equations) show that, at least for the marginal case of perfect fluid dynamics, the thermodynamic conditions can be interpreted. 

\subsection{Gradient theories of classical continua} There are several theoretical approaches for obtaining the weakly nonlocal evolution equations of classical continua. The thermodynamic analysis of Korteweg fluids has been the subject of research in various frameworks for over 50 years. Several methods have been reported, primarily fixing the entropy flux in advance, which requires the introduction of additional concepts like interstitial working (mentioned earlier), the balance of self-equilibrated forces, and multipolarity or virtual powers \cite{DunSer85a,GooCow72a,MehEta05a,FabEta11a,FabEta22a}. In this paper, a constructive, Second Law-based methodology was used without extra conditions beyond the extension of the CSS \cite{Van05a}. All previous analyzes obtain the same pressure of perfect Korteweg fluids as Sobrino [Eq.~\re{Sobp_eqn}] because the symmetry condition is not apparent without the Liu procedure. Remarkably, in Refs. \cite{CimEta09a}--\cite{GorEta21a}, the Liu procedure was applied to the analysis of various Korteweg fluid systems, but the holographic property was not recognized because the Liu equations could not be solved due to the particular treatment of the CSS \cite{Cim07a}. 

Our result regarding the perfect fluid is unique; there is no freedom to add an extra divergence term. The choice of parallel heat and entropy flux is a general aspect of non-equilibrium thermodynamics. Moreover, the dissipative part (the solution of the entropy inequality with linear constitutive equations) can be analyzed using the maximal entropy production method of Rajagopal \cite{Raj06a}. For Korteweg fluids, one can then obtain a rigorous solution \cite{HeiMal10a}. 

The gradient expansions of classical local equilibrium theories can be obtained without detailed thermodynamic analysis. However, Korteweg fluids are only one of the theories where thermodynamic conditions lead to significant improvement. This is also the case for the variational procedures of phase field theories \cite{HohHal77a,PenFif90a}. However, in the case of sophisticated static equilibrium (e.g., for generalized continua \cite{Eri99b}) or for distinguishing various dissipative effects in complicated geometries of non-Fourier heat conduction \cite{KovVan15a}, one cannot distinguish between the different theories without precise numerical calculations and dedicated experiments. The challenge is to control numerical dissipation and distinguish between numerical and physical dissipation \cite{PozsEta20a,TakEta22m}. Therefore, the direct connection of Korteweg fluids to multiple phenomena based on a large amount of experimental data is extremely helpful for benchmarking and testing the various methodologies.

{
	\subsection{Summary of discussion}

In this work the filling of the gaps between various branches of physics may shadow some novel aspects of the research. Therefore I highlight them in the following.

\begin{itemize}
	\item  A general methodology is presented to calculate the consequences of the entropy inequality for weakly nonlocal continua. The unified treatment of the divergence separation method of classical irreversible thermodynamics and Liu procedure of rational thermodynamics highlights their supplementary and constructive character. 
	\item The concept of classical holography in introduced in nonrelativistic field theories. 
	\item  Classical holographic property, the particular form of the potential field and the pressure tensor in ideal Fourier-Korteweg fluids is derived as the consequence of the Second Law of thermodynamics in the marginal case of zero dissipation.  
	\item  It was shown that the hydrodynamic form of quantum mechanics is not accidental. 
	\item  The general methodology, together with the presented conditions of continuum-quantum correspondence, represents a novel quantisation method. 
\end{itemize}
}

\section{Acknowledgments}   
This work was supported by a grant from the National Research, Development and Innovation Office (FK134277). The authors thank Robert Kovács, Mátyás Szűcs and Srbljub Simi\'c for valuable discussions. 

The research reported in this paper has been supported by the NRDI Fund (TKP2020 NC, Grant No. BME-NCS) based on the charter of bolster issued by the NRDI Office under the auspices of the Ministry for Innovation and Technology.

\bibliographystyle{unsrt}

\begin{thebibliography}{100}
\bibitem{ContraCov}
	The following is a simple explanation with relative quantities. The covector maps a vector to an invariant scalar. Let us assume that the transformation rule of a four-vector $A$ is Galilean; therefore, in a reference frame $K$, it has the time and spacelike components $A \overset{K}{\prec} (a,\pmb{a})$, and in the other reference frame $K'$, moving with a relative velocity $\pmb{v}$ related to $K$, the time and spacelike components are $A \overset{K'}{\prec} (a,\pmb{a}-a\pmb{v})$. The timelike part does not change, but the spacelike part is transformed via Galilean transformation. If the time and spacelike components of a four-covector $B$ in the reference frames $K$ and $K'$ are represented by $B \overset{K}{\prec} (b,\pmb{b})$ and $B \overset{K'}{\prec} (b',\pmb{b}')$, respectively, the transformation rules for the covector components follow from the invariance of $B\cdot A$: 
		$$
		B\cdot A= b' a'+ \pmb{b}'\cdot\pmb{a}' = b' a+ \pmb{b}'\cdot(\pmb{a}-a\pmb{v}) = (b'- \pmb{b}'\cdot \pmb{v})a+ \pmb{b}'\cdot\pmb{a} = b a+ \pmb{b}\cdot\pmb{a}.
		$$
	Therefore, $\pmb{b}' = \pmb{b}$, the spacelike part of the four-covector, does not transform, and the transformation rule of the timelike part is $b'= b+\pmb{b}\cdot\pmb{v}$.
\bibitem{aNoll} 	
	{Remarkably, Noll himself joined the critics of Noll objectivity to complain about the insufficient formulation and argued that the concept must be further developed; the Descartes product of space and time is not a correct representation of spacetime in nonrelativistic physics \cite{Nol04m,Nol06a,NolSeg10a}.}
\end{thebibliography}

\end{document}